\documentclass{sig-alternate}
\usepackage{algpseudocode,subfig}
\usepackage{algorithm}
\usepackage{lipsum}
\usepackage{graphicx}
\algrenewcommand\algorithmicindent{1.0em}
\begin{document}

\title{Evaluating the Potential of a Dual Randomized Kaczmarz Solver for Laplacian Linear Systems}

\numberofauthors{3}
\author{
\alignauthor
Erik G. Boman\footnotemark[1]\\
       \affaddr{Sandia National Laboratories}\\
       \affaddr{Center for Computing Research}\\
       \email{egboman@sandia.gov}
\alignauthor
Kevin Deweese\footnotemark[2]\\
\affaddr{UC Santa Barbara}\\       
\affaddr{Department of Computer Science}\\
       \email{kdeweese@cs.ucsb.edu}
\alignauthor 
John R. Gilbert\footnotemark[2]\\
\affaddr{UC Santa Barbara}\\
\affaddr{Department of Computer Science}\\
\email{gilbert@cs.ucsb.edu}
}

\maketitle
\footnotetext[1]{Sandia is a multi-program laboratory managed and
operated by Sandia Corporation,
a wholly owned subsidiary of Lockheed Martin Corporation,
for the U.S.
Department of Energy's National Nuclear Security Administration under contract
DE-AC04-94AL85000.}
\footnotetext[2]{Supported by Contract \#618442525-57661 from Intel Corp.
and Contract \#DE-AC02-05CH11231 from DOE Office of Science.}
\begin{abstract}
A new method for solving
Laplacian linear systems proposed by Kelner et al.\ involves the
random sampling and update of fundamental cycles
in a graph.
Kelner et al.\ proved asymptotic bounds on the complexity of
this method but did not report experimental results.
We seek to both evaluate the performance of this
approach and to explore improvements to it in practice.
We compare the performance of this method to other Laplacian
solvers
on a variety of real
world graphs.
We consider different ways to improve
the performance of this method by exploring
different ways of choosing the set of cycles and
the sequence of updates, with
the goal of providing more flexibility and potential parallelism.
We propose a parallel
model of the Kelner et al.\ method, for evaluating
potential parallelism in terms of the span of edges updated at
each iteration.
We provide experimental results comparing the potential parallelism
of the fundamental cycle basis and our extended cycle set.
Our preliminary experiments show that choosing a non-fundamental
set of cycles
can save significant work compared to a fundamental cycle basis.
\end{abstract}




\section{Introduction}
\subsection{Graph Laplacians}
The Laplacian matrix of a weighted, undirected graph is defined as $L=D-A$,
where $D$ is the diagonal matrix containing the sum of
incident edge weights and
$A$ is the weighted adjacency matrix. The Laplacian is symmetric and positive
semidefinite. The Laplacian is also defined for directed graphs
\cite{AgaevChebotarev2005}. Because they are not symmetric,
most of the solvers discussed
in this paper do not apply, and efficient solution techniques remain
an open problem.
Solving linear systems on the Laplacians of structured graphs,
such as two and three dimensional meshes, has long been important in
finite element analysis (with applications in electrical and thermal
conductivity, and fluid flow modeling \cite{BHV2008})
and image processing (with applications in image segmentation,
inpainting, regression, and classification
\cite{KMT2011,McCannPollard2008,Grady2006}).
More recently,
solving linear systems on the graph Laplacians of
large graphs, with irregular degree distributions,
has emerged as an important computational task in network analysis
(with applications to maximum flow problems \cite{CKMST2011},
graph sparsification \cite{SpielmanSrivastava2011},
and spectral clustering \cite{KhoaChawla2015}).

Most applied work on Laplacian solvers has been on
preconditioned conjugate gradient (PCG) solvers, including support graph
preconditioners \cite{GrembanPHD1996,BGHNT2006,BCHT2004},
or on specialized multigrid
methods \cite{LivneBrandt2012,KMT2011}.
Several of these methods seem efficient in practice, but
none have asymptotic performance guarantees based on the size
of the graph.

The theoretical computer science community has developed several methods,
which we refer to generally as combinatorial Laplacian solvers.
These solvers have good complexity bounds but, in most cases,
no reported experimental results.
Spielman and Teng
\cite{SpielmanTeng2004} first showed
how to solve these problems in
linear times polylogarithmic work, later
improved upon by Koutis, Miller, and Peng \cite{KMP2010},
but their algorithms
do not yet have a practical implementation.
An algorithm proposed by Kelner et al.\ \cite{KOSZ2013}
has the potential to solve these linear
systems in linear times polylogarithmic
work with a simple, implementable algorithm.

\subsection{The Dual Randomized Kaczmarz Algorithm}
The inspiration for the algorithm
proposed by Kelner et al.\ \cite{KOSZ2013},
which we refer to as Dual Randomized Kaczmarz (DRK),
is to treat graphs as electrical
networks with resistors on the edges.
For each edge, the weight is the inverse of the
resistance. We can think of vertices as having an electrical potential
and a net
current at every vertex, and define vectors of these potentials and currents as
$v$ and $f$ respectively. These vectors are related by the
linear
system $L v = f$. Solving this system is equivalent to finding
the set of voltages that satisfies the net ``injected'' currents.
Kelner et al.'s DRK
algorithm solves this problem with an optimization algorithm in the dual space,
which finds the optimal currents on all of the edges subject to the constraint
of zero net voltage around all cycles. They use Kaczmarz projections
\cite{Kaczmarz1937} to adjust currents on one cycle at a time,
iterating until convergence.

We will also refer to the
Primal Randomized Kaczmarz (PRK) method that applies Kaczmarz projections
in the primal space \cite{StrohmerVershynin2009}. One sweep of PRK performs
a Kaczmarz projection with every row of the matrix. Rows are taken in random
order at every sweep.

DRK iterates over a set of \textit{fundamental cycles},
cycles formed by adding
individual edges to a spanning tree $T$.
The fundamental cycles are a basis for the space of all cycles in the graph
\cite{Diestel2010}.
For each off-tree edge $e$, we define
the resistance $R_e$ of the cycle $C_e$ that is
formed by adding edge $e$ to the spanning tree
as the sum of the
resistances around the cycle,
\[
R_e = \sum_{e' \in C_e} r_{e'}
\]
which is thought of as approximating the
resistance of the off-tree edge $r_e$. DRK
chooses cycles randomly,
with probability proportional to $R_e/r_e$.

The performance of the algorithm depends on the sum of
these approximation ratios, a property of the spanning tree
called the \textit{tree condition number}
\[
\tau(T) = \sum_{e \in E \setminus T} \frac{R_e}{r_e}.
\]
The number of iterations of DRK is proportional to the
tree condition number.
Kelner et al.\ use
a particular type of spanning tree with low tree condition
number, called a \textit{low stretch tree}. Specifically, they
use the
one described by Abraham and Neiman \cite{AbrahamNeiman2012}
with $\tau=O(m \log{n} \log{\log{n}})$, where $n$ refers
to the number of vertices and $m$ refers to the number of edges
of the original graph.
The work of one iteration is naively the cycle length, but
can be reduced to $O(\log{n})$ with a fast data structure,
yielding $O(m \log{n}^2 \log{\log{n}})$ total work.

\subsection{Overview}
The rest of the paper is organized as follows.
In Section 2 we survey  the related experimental work.
Section 3 is an initial evaluation of the DRK algorithm
as compared to PCG and PRK.
We present our new ideas for improving the performance of DRK in Section 4.
In this section we also consider how to perform cycle updates
in parallel.
Section 5 is an evaluation of the new ideas proposed in Section 4.

\subsection{Related Experimental Work}
As the DRK algorithm is a recent and theoretical
result, there are few existing implementations or performance results.
Hoske et al.\ implemented the DRK algorithm in C++ and did timing comparisons
against unpreconditioned CG on two sets of generated graphs \cite{HLMW15}.
They concluded that the solve time of DRK
does scale nearly linearly.
However, several factors make the running time too large in practice,
including large tree stretch and cycle updates
with unfavorable memory access patterns.
They cite
experimental results by Papp \cite{Paap2014}, which
suggest that the theoretically low stretch tree algorithms
are not significantly better than min-weight spanning trees in
practice, at least on relatively small graphs.

Chen and Toledo \cite{ChenToledo2003} experimented with
an early and somewhat different combinatorial approach to Laplacians
called support graph preconditioners. They demonstrated that support graph
preconditioners can outperform incomplete Cholesky on certain problems.
There has also been some experimental work in implementing the local
clustering phase of the Spielman and Teng algorithm \cite{ZLM2013}.

\section[Initial Evaluation of DRK and Comparison to PCG and PRK]
{Initial Evaluation of DRK\\and Comparison to PCG and PRK}

\subsection{Experimental Design}
Our initial study of DRK measures performance in terms of
work instead of time, and uses a somewhat more diverse graph test set
than Hose et al.\ \cite{HLMW15}.
We implemented the algorithm in Python with Cython to see how it compared
against PCG (preconditioned with Jacobi diagonal scaling) and PRK.
However, we did
not implement a low stretch spanning tree.
Instead we use a low stretch heuristic that ranks and greedily
selects edges by the sum
of their incident vertex degrees (a cheap notion of centrality).
We have found that this works well on unweighted graphs.
We also did not implement
the fast data structure Kelner et al.\ use to update cycles in $O(\log{n})$
work.

\begin{figure*}[htb!]
\centering
\subfloat[Mesh-like Graphs]{
\includegraphics[scale=.35]{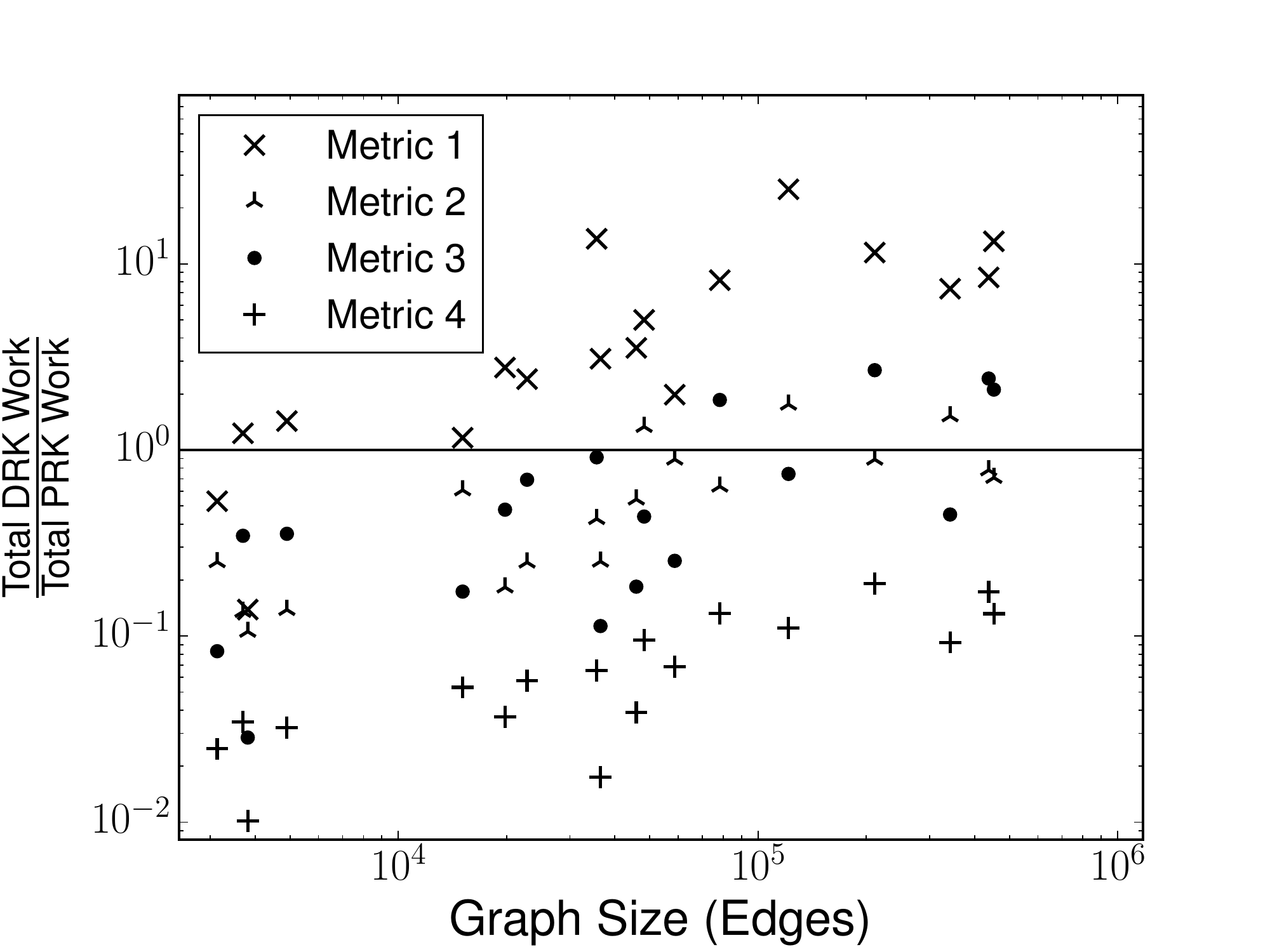}}
\subfloat[Irregular Graphs]{
\includegraphics[scale=.35]{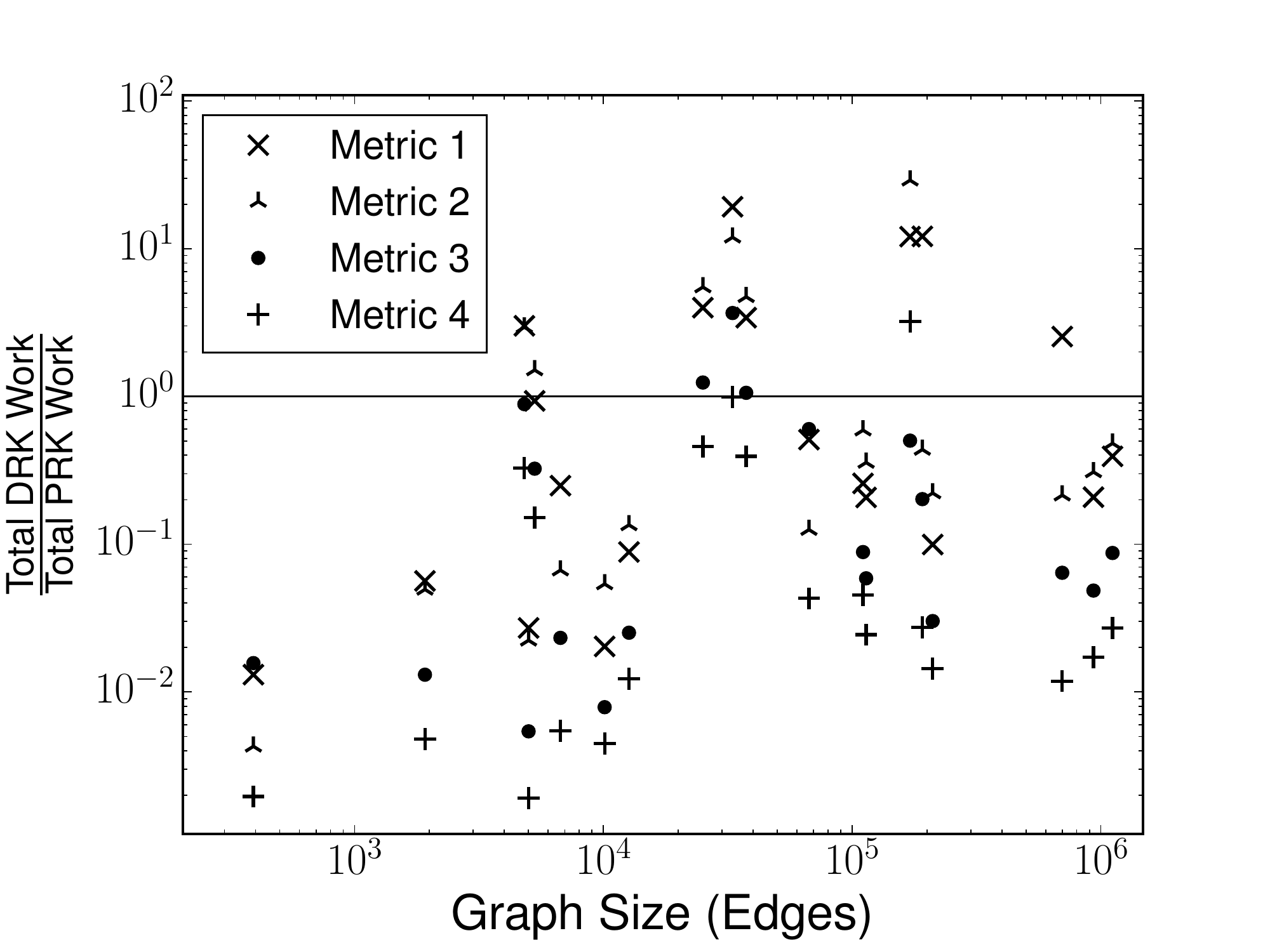}}
\caption{DRK vs. PRK: Relative work of DRK to PRK work under the four cost
metrics is shown (PRK is better than DRK at points above the line.)\label{fig:prkworkcompare}}
\end{figure*}

\begin{figure*}[htb!]
\centering
\subfloat[Mesh-like Graphs]{\includegraphics[scale=.35]{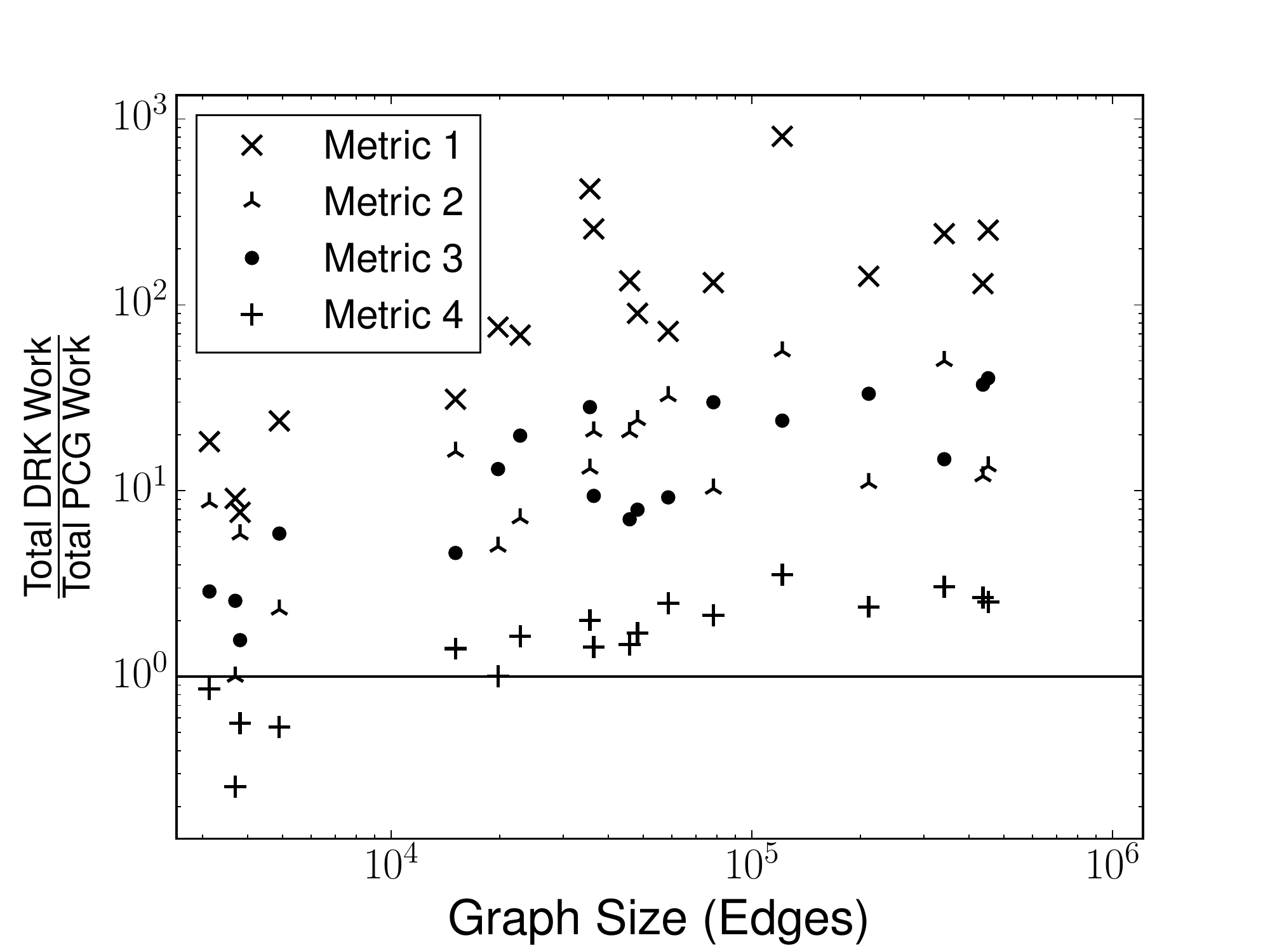}}
\subfloat[Irregular Graphs]{\includegraphics[scale=.35]{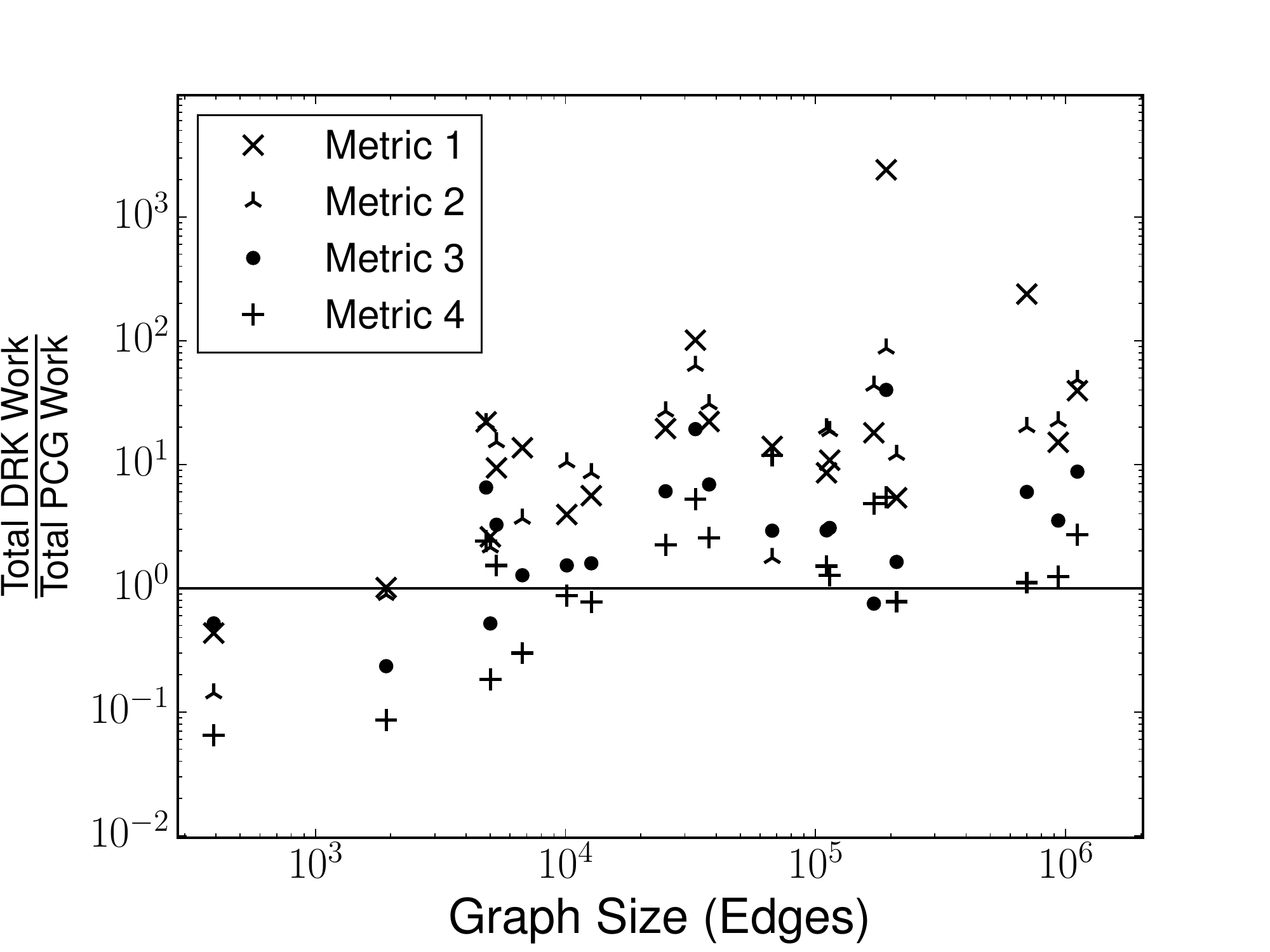}}
\caption{DRK vs. PCG: Relative work of DRK to PCG work under the four cost metrics
is shown
(PCG is better than DRK at points above the line.)\label{fig:pcgworkcompare}}
\end{figure*}

\begin{figure*}[htb!]
\centering
\subfloat[Mesh-like Graphs]{\includegraphics[scale=.35]{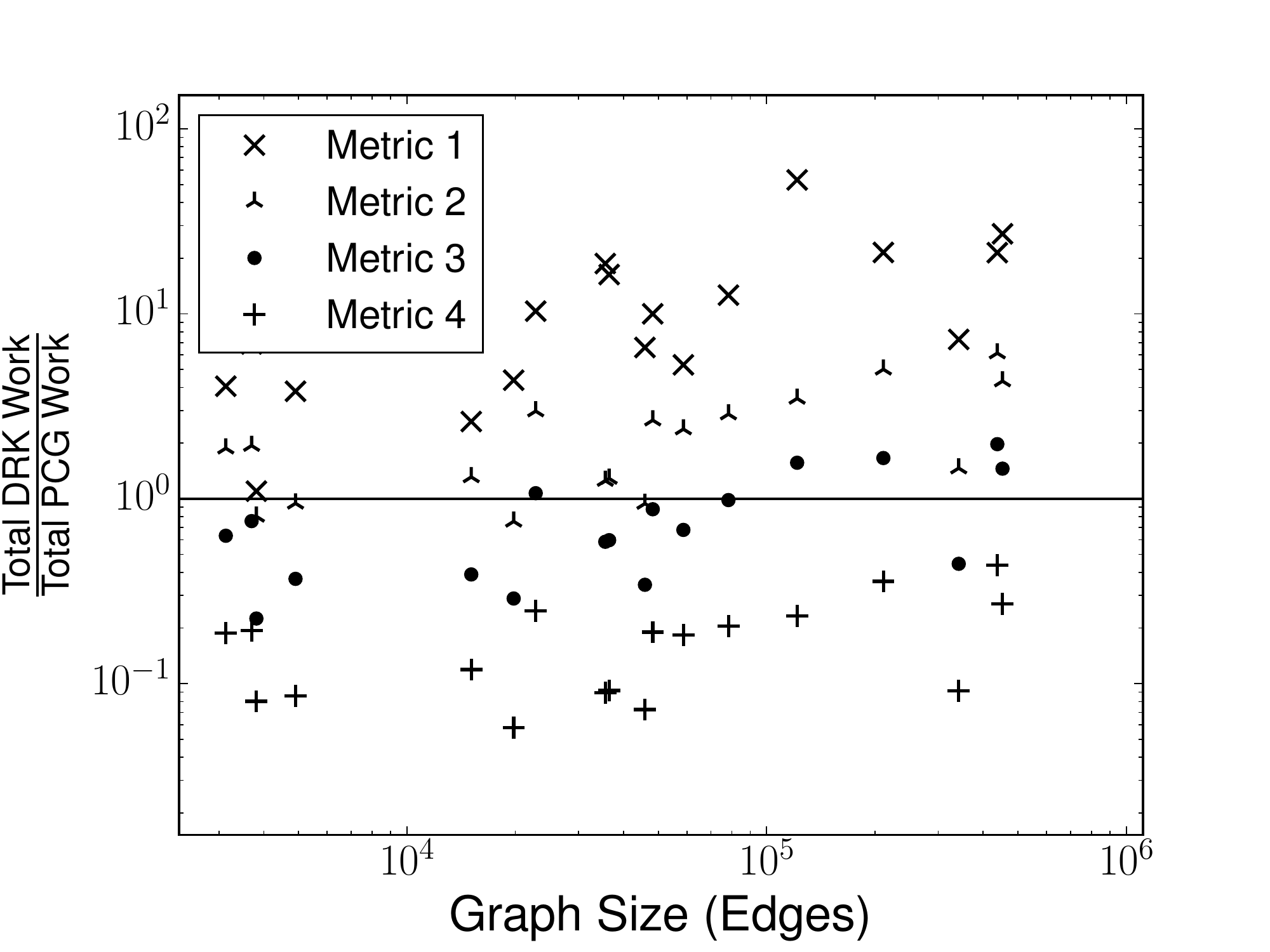}}
\subfloat[Irregular Graphs]{\includegraphics[scale=.35]{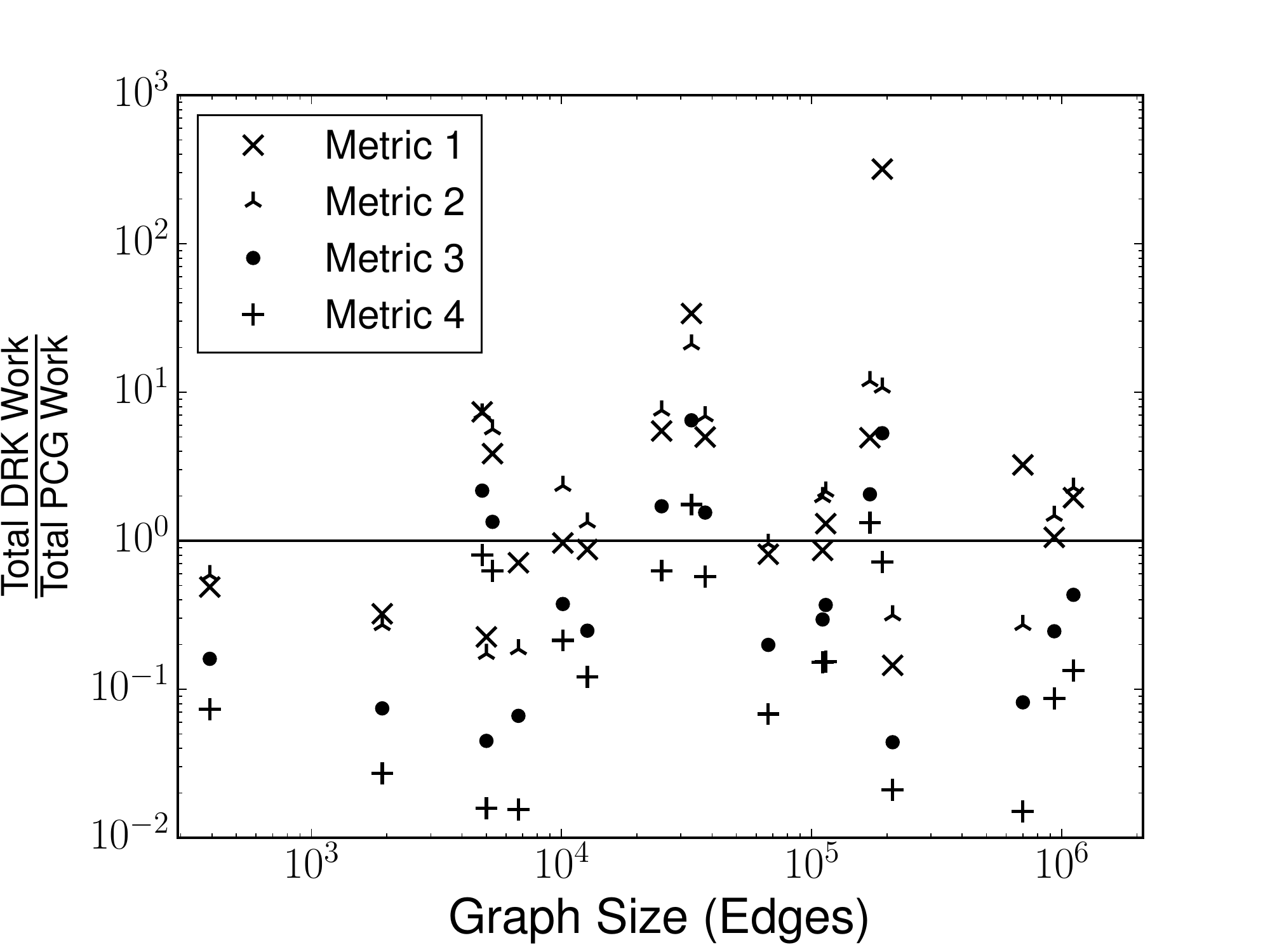}}
\caption{DRK vs. PCG Converged to Actual Error: Relative work of DRK to PCG work under the four cost metrics is \
shown, convergence tolerance is norm of actual error within $10^{-3}$\label{fig:pcgactualerr}}
\end{figure*}

We do not report
wall clock time, since our DRK implementation is not highly optimized.
Instead we are interested in measuring the total work. For PCG the work is
the number of nonzeros in the matrix for every iteration, plus the work
of applying the preconditioner at every iteration
(number of vertices for Jacobi).
For PRK the work is the number of nonzero
entries of the matrix for every sweep,
where a sweep is a Kaczmarz projection against all the rows of $L$.
As the DRK work will depend on data structures and implementation,
we consider four different costs for estimating the work
of updating a single cycle, which we refer to as cost metrics.
The first metric is the cost of updating
every edge in a cycle, which is included
because it is the naive implementation
we are currently using. The second metric relies upon
the data structure described by Kelner et al., which can
update the fundamental cycles
in $O(\log{n})$ work.
This may be an overestimate when the cycle length is actually less than
$\log{n}$.
The third metric considers a
hypothetical $\log$ of cycle length update method
which we do not know to exist, but is included as a hopeful
estimate of a potentially better update data structure.
The last metric costs $O(1)$ work per cycle, which is
included because we surely cannot do better than this.
\begin{description}
\item[Metric 1.] $\text{cycle length}$ (naive)
\item[Metric 2.] $\log{n}$ (using fast update data structure)
\item[Metric 3.] $\log(\text{cycle length})$ (optimistic)
\item[Metric 4.] $1$ (lower bound)
\end{description}

We ran experiments on all the mesh-like graphs
and irregular graphs shown in Appendix Table \ref{table:graphsizes}.
Mesh-like graphs come from more traditional applications
such as model reduction and
structure simulation, and contain a more regular degree distribution.
Irregular graphs come from electrical, road, and social networks, and
contain a more irregular, sometimes exponential, degree distribution.
Most of these graphs are in the University of Florida (UF)
sparse matrix collection \cite{DavisHu2011}.
We added a few 2D and 3D grids along with a few graphs generated with
the BTER generator \cite{KPPS2014}.
We removed weights and in a few cases symmetricized the matrices by
adding the transpose.
We pruned the graphs to the largest connected component of
their 2-core, by successively removing
all degree 1 vertices, since DRK operates
on the cycle space of the graph.
The difference between the original graph and the 2-core
is trees that are pendant on the original graph. These can be solved in linear
time so we disregard them to see how solvers compare on just
the structurally interesting
part of the graph.

We solve to a relative residual tolerance of $10^{-3}$.
The Laplacian matrix is singular with a nullspace dimension of one
(because the pruned graph is connected).
For DRK and PRK this is not a problem, but for PCG we must handle
the non-uniqueness of the solution.
We choose to do this
by removing the last row and column of the matrix.
We could also choose to orthogonalize the solution against
the nullspace inside the algorithm, but
in out experience the performance results are similar.

We also
ran a set of PCG vs.\ DRK experiments where the convergence criteria
is the actual error within $10^{-3}$.
We traded off accuracy in the solution to run more
experiments and on larger graphs.
We do this knowing the solution in advance.
One of the interesting results
of the DRK algorithm is that, unlike PCG and PRK,
convergence does not depend on the condition
number of the matrix, but instead just on the tree condition number.
Since higher condition number can make small residuals
less trustworthy, we wondered whether convergence in
the actual error yields different results.

\subsection{Experimental Results} 
We compare DRK to the other solvers by examining the ratio
of DRK work to the work of the other solvers.
The ratio of DRK work to PRK
work is plotted in Figure \ref{fig:prkworkcompare},
separated by graph type.
Each vertical set of four points are results for
a single graph, and are sorted on the x axis by graph size.
The four points represent the ratio
of DRK work to PRK work under all four cost metrics.
Points above the line indicate DRK performed more work while
points below the line indicate DRK performed less work.
Similar results for the PCG comparison are shown
in Figure \ref{fig:pcgworkcompare}.
Another set of PCG comparisons, converged to the actual
error, is shown in Figure \ref{fig:pcgactualerr}.

An example
of the convergence behavior on the USpowerGrid graph is shown
in Figure \ref{fig:actualerrexample}. This plot indicates
how both the actual error and relative error behave during
the solve for both PCG and DRK.
A steeper slope indicates faster convergence.
Note this only shows metric 1 work for DRK.

\subsection{Experimental Analysis}
In the comparison to PRK (shown in Figure \ref{fig:prkworkcompare}),
DRK is often better with cost metrics 3 and 4.
On a few graphs, mostly in the irregular
category, DRK outperforms PRK in all cost metrics
(all the points are below the line.)
In the comparison to PCG (shown in Figure \ref{fig:pcgworkcompare}),
DRK fares slightly better for the irregular graphs,
but on both graph sets
these results are somewhat
less than promising.
PCG often does better (most of the points are above the line).
Even if we assume unit cost for cycle updates, PCG outperforms DRK.
The performance ratios also seem to get worse as graphs get larger.

\begin{figure}[htb!]
\centering
\includegraphics[scale=.4]{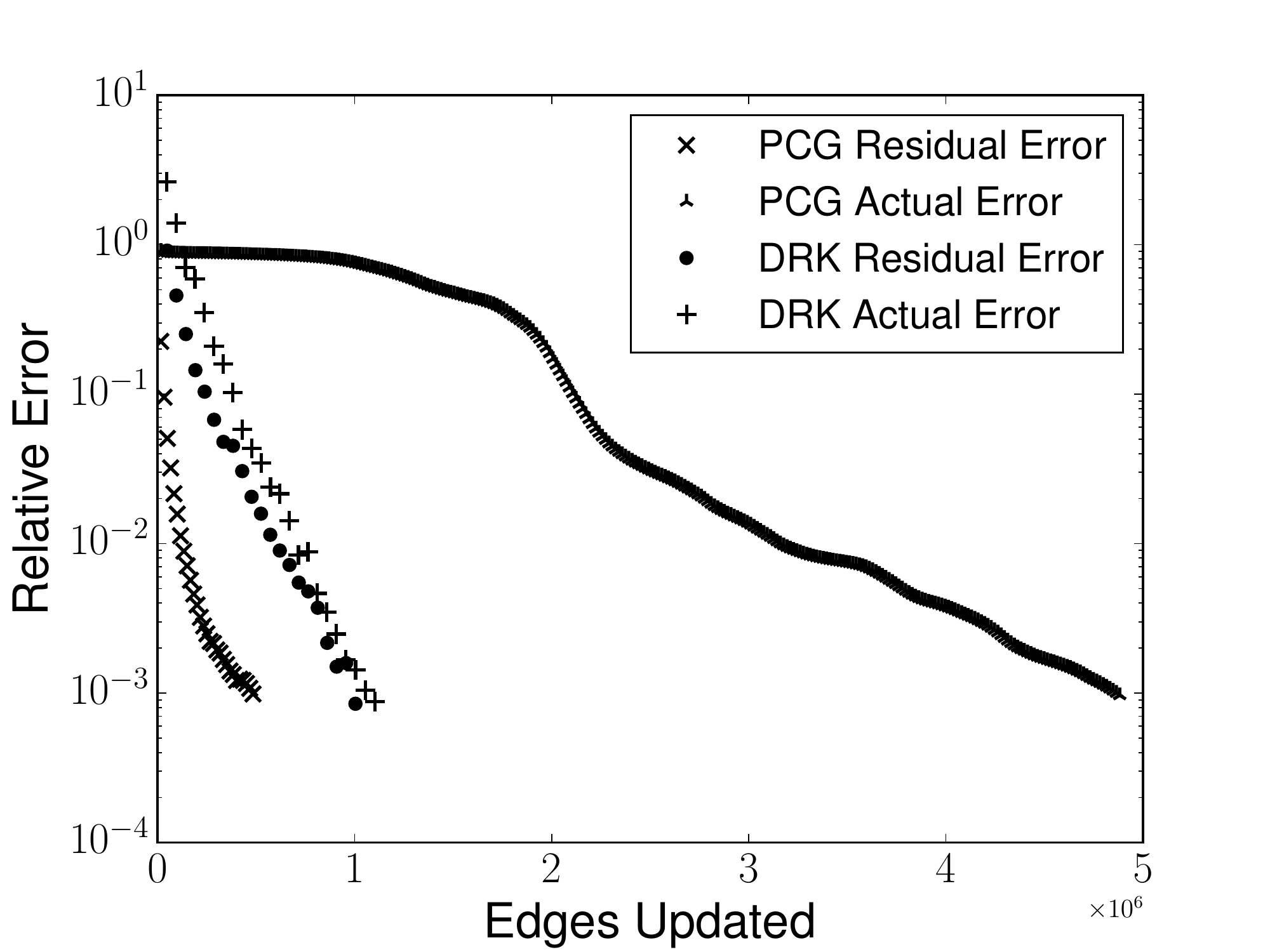}
\caption{DRK and PCG Convergence Behavior on USpowerGrid: Relative
residual error and actual error are shown for both solvers over the iterations
required for convergence.
\label{fig:actualerrexample}}
\end{figure}

The results concerning the actual error
(shown in Figure \ref{fig:pcgactualerr})
are very interesting as
they are quite different than those with the residual tolerance. For all of the
mesh graphs, considering the actual error makes DRK look more promising.
The relative performance of cost metrics 3 and 4 are now typically better for
DRK than PCG. However, PCG is still consistently better with cost metrics
1 and 2.
For some of the irregular
graphs, the convergence behavior is similar,
but for others things look much better
when considering actual error. Informally the number of edges
updated by DRK did
not change much when switching convergence criteria, but PCG work often
increased.
The USpowerGrid example (shown in Figure \ref{fig:actualerrexample})
gives a sense of this. The residual error and actual
error decrease similarly for DRK, but the actual
error curve for PCG decreases much
more slowly for the actual error.

\section{New Algorithmic Ideas}
We consider ways in which DRK could be improved by
altering the choice of cycles and their updates.
Our goals are both to reduce total work and to identity
potential parallelism in DRK. To this end
we are interested in measuring the number of
\textit{parallel steps}, the longest number of steps a single
thread would have to perform before before convergence,
maximized over all threads.
Parallel steps are measured in terms of the four cost metrics
described in section 2.
We will also define
the \textit{span} \cite{CLRS2009},
or critical path length, which is the number
of parallel steps with unbounded threads.

\subsection{Expanding the Set of Cycles}
Sampling fundamental cycles with respect to a tree may require updating several
long cycles which will not be edge-disjoint. It would be preferable to update
edge-disjoint cycles, as these updates could be done in
parallel.
The cycle set we use does not need to be a basis, but it does need to
span the cycle space.
In addition to using a cycle basis from a spanning tree, we will use
several small, edge-disjoint cycles. We expect that having threads update these
small cycles is preferable to having them stand idle.

\subsubsection{2D Grid Example}
\begin{figure}[htb]
\centering
\begin{minipage}{.22\textwidth}
\subfloat[]{
\includegraphics[width=3.7cm]{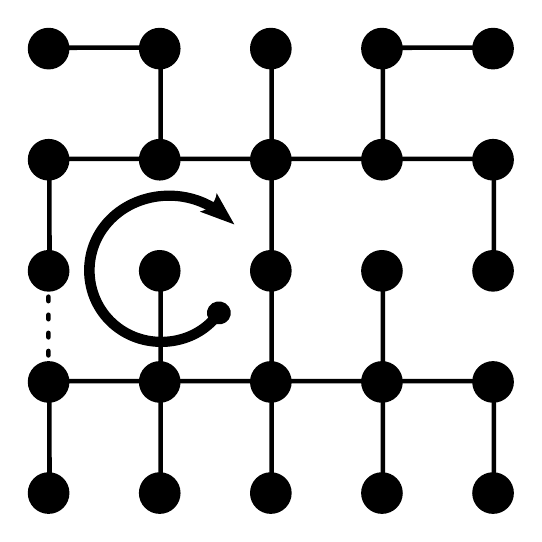}
}
\end{minipage}
\begin{minipage}{.22\textwidth}
\subfloat[]{
\includegraphics[width=3.7cm]{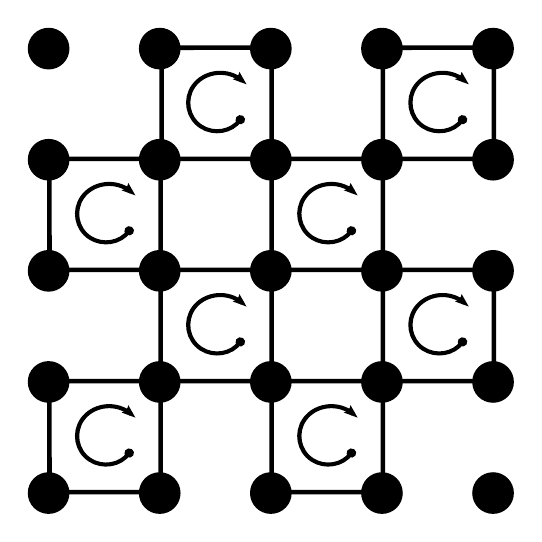}
}
\end{minipage}
\begin{minipage}{.22\textwidth}
\subfloat[]{
\includegraphics[width=3.7cm]{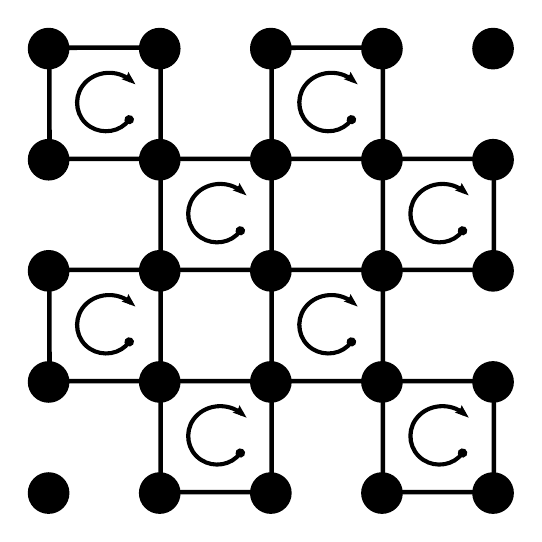}
}
\end{minipage}
\begin{minipage}{.22\textwidth}
\subfloat[]{
\includegraphics[width=3.7cm]{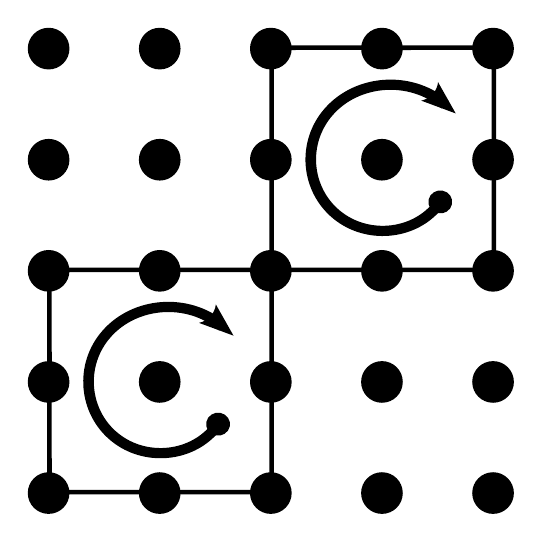}
}
\end{minipage}
\caption{Grid Cycles: (a) Fundamental cycles are formed by adding edges to the spanning
tree. (b-c) First level facial cycles are shown, grouped into edge-disjoint sets. (d)
Second level facial cycles are formed by adding smaller facial cycles.}
\label{grid}
\end{figure}

A simple example of a different cycle basis is the
2D grid graph, shown in Figure \ref{grid}.
In the original DRK, cycles are selected by adding
off-tree edges to the spanning tree as in Figure \ref{grid}(a).
As the 2D grid graph is planar, the faces of
the grid are the regions bounded by edges, and we refer
to the cycles that enclose these regions as \textit{facial cycles}.
We consider using these cycles to perform updates of DRK, as
the facial cycles span the
cycle space of a planar graph \cite{Diestel2010}.
Half of these cycles
can be updated at one iteration and then the other half can be updated
during the next iteration, in a checkerboard fashion, as in
Figures \ref{grid}(b)(c). Furthermore, to speed up convergence, smaller cycles
can be added together to form larger cycles (in a multilevel fashion)
as in Figure \ref{grid}(d).

\begin{figure}[htb!]
\centering
\includegraphics[scale=.4]{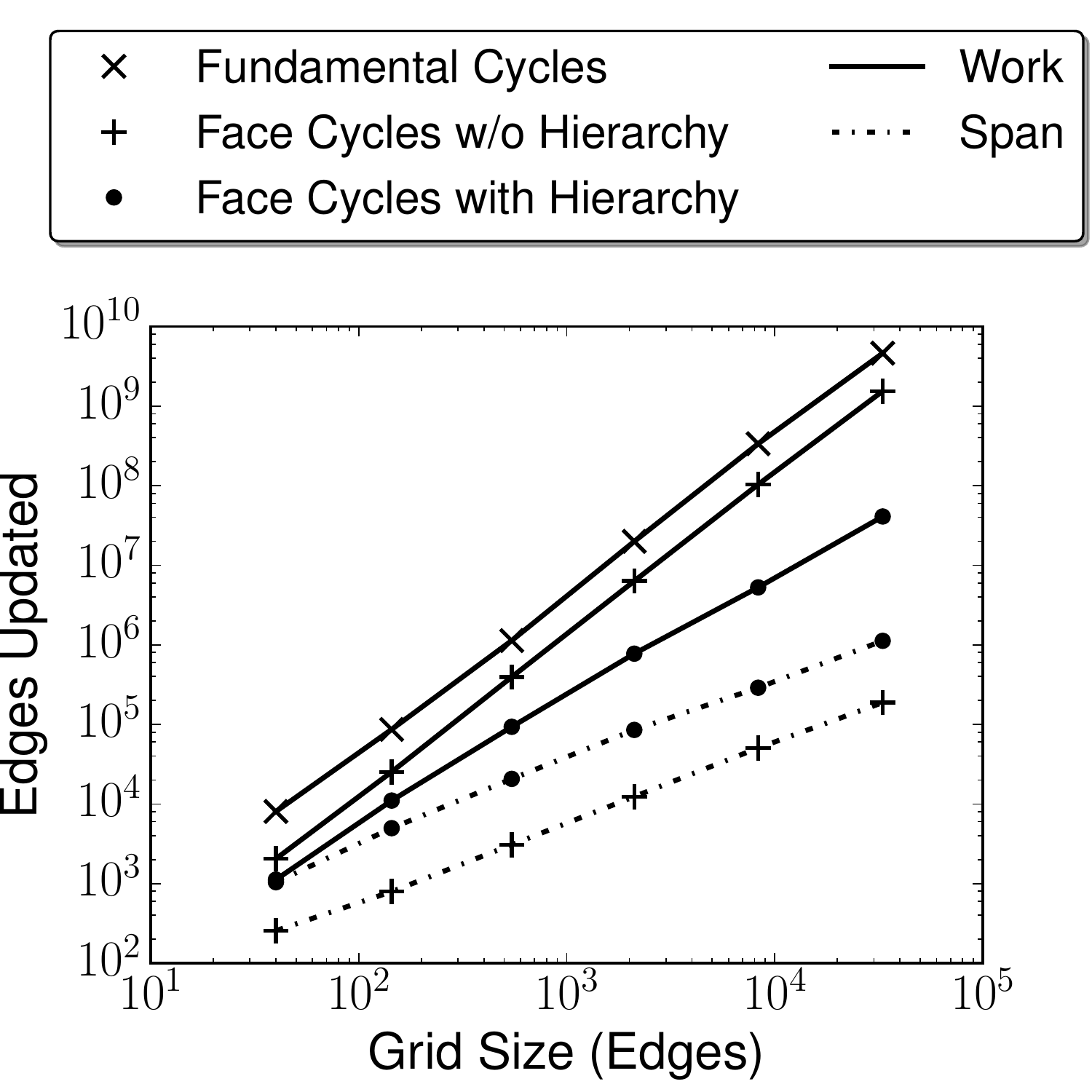}
\caption{Grid Cycle Performance:
Work and span of DRK using facial cycles and fundamental cycles
for two dimensional grids of various sizes.}
\label{plot:gridresults}
\end{figure}

We implemented such a cycle update scheme using the grid
facial cycles, and performed experiments to see how the facial cycles affected
the total work measured in both the number of cycles updated (metric 4)
and edges updated (metric 1).
With the facial cycles,
the span per iteration is the cost of updating two cycles at each level.
We ran experiments with
and without the hierarchical combination of the facial cycles against the original
set of fundamental cycles.
In the case of the fundamental cycles we use H trees \cite{AKPW1995}, which
have optimal stretch $O(\log{n})$.
Solutions were calculated to a residual tolerance of $10^{-6}$.
The accuracy here is slightly better than the rest of the experiments
since these experiments were faster.

The results shown in Figure \ref{plot:gridresults}
indicate that the facial cycles
improve both the work and span. Using a hierarchical update scheme reduces
the total number of edges updated. However as this requires updating larger
cycles it has a worse span than simply using the lowest level of cycles.

\subsubsection{Extension to General Graphs}

We refer to small cycles we add
to the basis as local greedy cycles. Pseudocode for finding
these cycles is shown in Algorithm 1.
We construct this cycle set by attempting to find a small cycle containing
each edge
using a truncated breadth-first search (BFS).
Starting with all edges unmarked, the algorithm selects an unmarked edge and attempts
to find a path between its endpoints. This search is truncated by bounding the
number of edges searched so that each search
is constant work and constructing the entire set is $O(m)$ work.
If found, this path plus the edge forms a cycle, which
is added to the new cycle set, and all edges used are marked.
Table \ref{table:graphsizes} shows the number of local greedy cycles
found for all the test graphs when the truncated BFS was
allowed to search 20 edges.
Greedy cycles were found in all the graphs except for tube1,
all of whose vertices had such high degree that
searching 20 edges was not enough to find a cycle.

\begin{algorithm}[]
\begin{algorithmic}[]
\Function  {Local-Greedy}{G}
\For {$e_{i,j} \in E$}
\If {$e_{i,j}$ unmarked}
\State $p_{i,j} = \textrm{Truncated-BFS}(G\setminus(e_{i,j}),i,j,max\_edges)$
\State Add $p_{i,j} + e_{i,j}$ to cycle set
\State Mark all edges in $p_{i,j}+e_{i,j}$
\EndIf
\EndFor
\EndFunction
\end{algorithmic}
\label{alg:LocalGreedy}
\caption{Local Greedy Finder}
\end{algorithm}

Adding additional cycles to the cycle basis means we need
new probabilities with which to sample all the cycles. Since in the unweighted
case, the stretch of a cycle is just its total length,
it seems natural to update cycles proportional
to their length.

\begin{figure}[h]
\centering
\minipage{.25\textwidth}
\includegraphics[scale=.5]{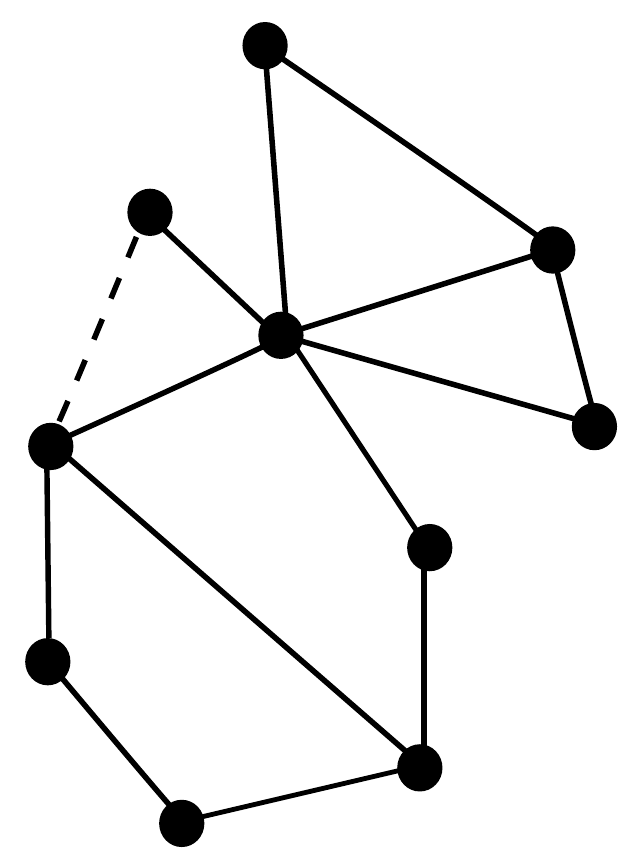}
\endminipage
\minipage{.25\textwidth}
\includegraphics[scale=.5]{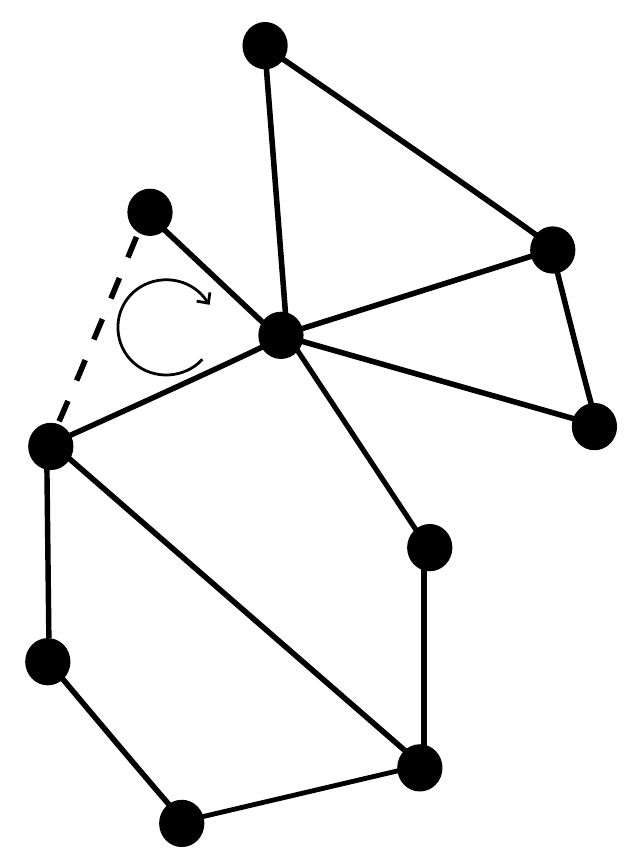}
\endminipage
\caption{Local Greedy Cycles: An edge is selected on the left and a local
greedy search is performed to find the cycle on the right.}
\label{fig:localgreedy}
\end{figure}

\subsection{Cycle Sampling and Updating in Parallel}
\begin{figure}[h]
\centering
\includegraphics[scale=.6]{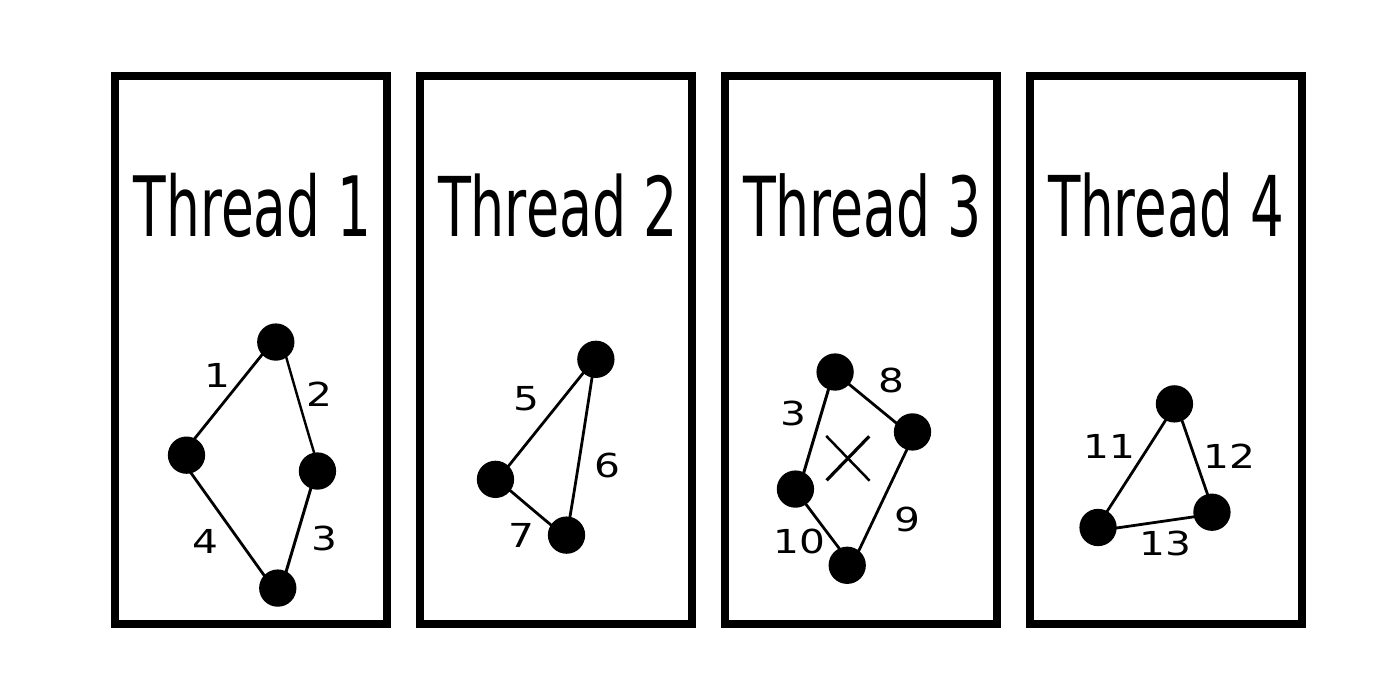}
\caption{Example of Processors Selecting Cycles: Threads 1, 2, and 4 select
edge-disjoint cycles, but thread 3 selects a cycle with edge 3 already in use.
Thread 3 will go idle for an iteration.}
\label{parallelcycle}
\end{figure}

In the original DRK algorithm, cycles are chosen one at a time with
probability proportional to stretch.
We propose a parallel update scheme
in which multiple threads each select a cycle,
at every iteration, with
probability proportional to cycle length. If two threads
select cycles that share an edge, one of the threads goes idle for that
iteration. In Figure \ref{parallelcycle}, threads 1, 2, and 4 select
edge-disjoint cycles.
However the third processor selects a cycle which contains edge 3,
which is already in use by the cycle on thread 1. Processor 3 sits this
iteration out while the other processors update their cycles.

We compute several measures of parallel performance.
The first is simply the number of iterations. The second is the total work
across all threads at every iteration. Lastly we measure the span,
or critical path length. This is the maximum of the work over all threads,
summed over all the iterations.

We envision threads working in a shared memory environment on a graph
that fits in memory.
This might not be realistic in practice as there must be some communication of
which edges have already been used which might be too expensive relative to the
cost of a cycle
update. However we are simply interested in measuring
the potential parallelism, thus we ignore any communication cost.

The parallel selection scheme conditions
the probabilities with which cycles are selected on
edges being available
\[
p(C_e) = \frac{1}{\tau}\frac{R_e}{r_e} p(e' \in C_e \text{ available}).
\]
This scheme creates a bias
towards smaller cycles with less conflicting
edges as more threads are added, which can increase total work.

\newpage
\section{Experiments and Results}
\subsection{Experimental Design}
We performed experiments on a variety of unweighted graphs
from the UF Sparse Matrix Collection
(the same set as in Section 2,
shown in Table \ref{table:graphsizes}).
Again we distinguish between mesh-like graphs and irregular graphs.
We also use a small test set for weak scaling experiments, consisting
of 2D grids and BTER graphs.

We continue to use our Python/Cython implementation of DRK,
without a guaranteed low stretch spanning
tree or a cycle update data structure.
The code does not run in parallel,
but we simulate parallelism on multiple threads by selecting and updating
edge-disjoint cycles at every iteration as described above.

Our experiments consist of two sets of strong scaling experiments,
the spanning tree cycles with and without local greedy cycles, up to
32 threads. We set a relative residual tolerance of $10^{-3}$.
Again we sacrifice accuracy to run more experiments on larger graphs.
We consider the same four cycle update cost metrics
as in Section 2: \text{cycle length}, $\log{n}$,
$\log(\text{cycle length})$, and unit cost update.
However in the case of the local greedy cycles, which cannot use the
$\log{n}$ update data structure, we always just
charge the number of edges in a cycle.
For all the cost models, we measure the total work required for convergence
and the number of parallel steps taken to converge. For metric 4 these
will be the same. A condensed subset of the scaling results
is shown in Table \ref{table:results}.

\subsection{Experimental Results}
\begin{figure*}[htb]
\centering
\subfloat[Metric 1]{\includegraphics[scale=.35]{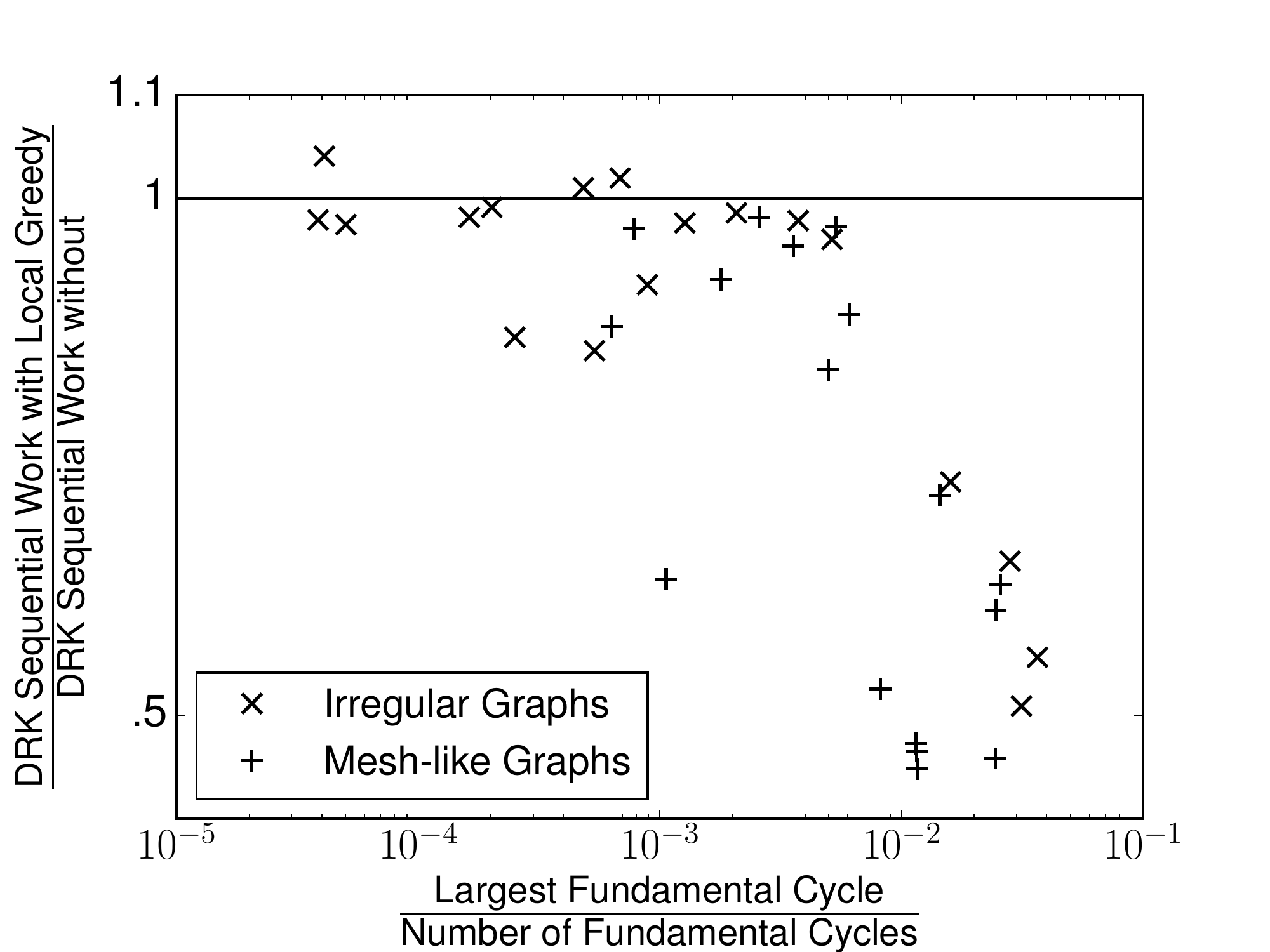}}
\subfloat[Metric 4]{\includegraphics[scale=.35]{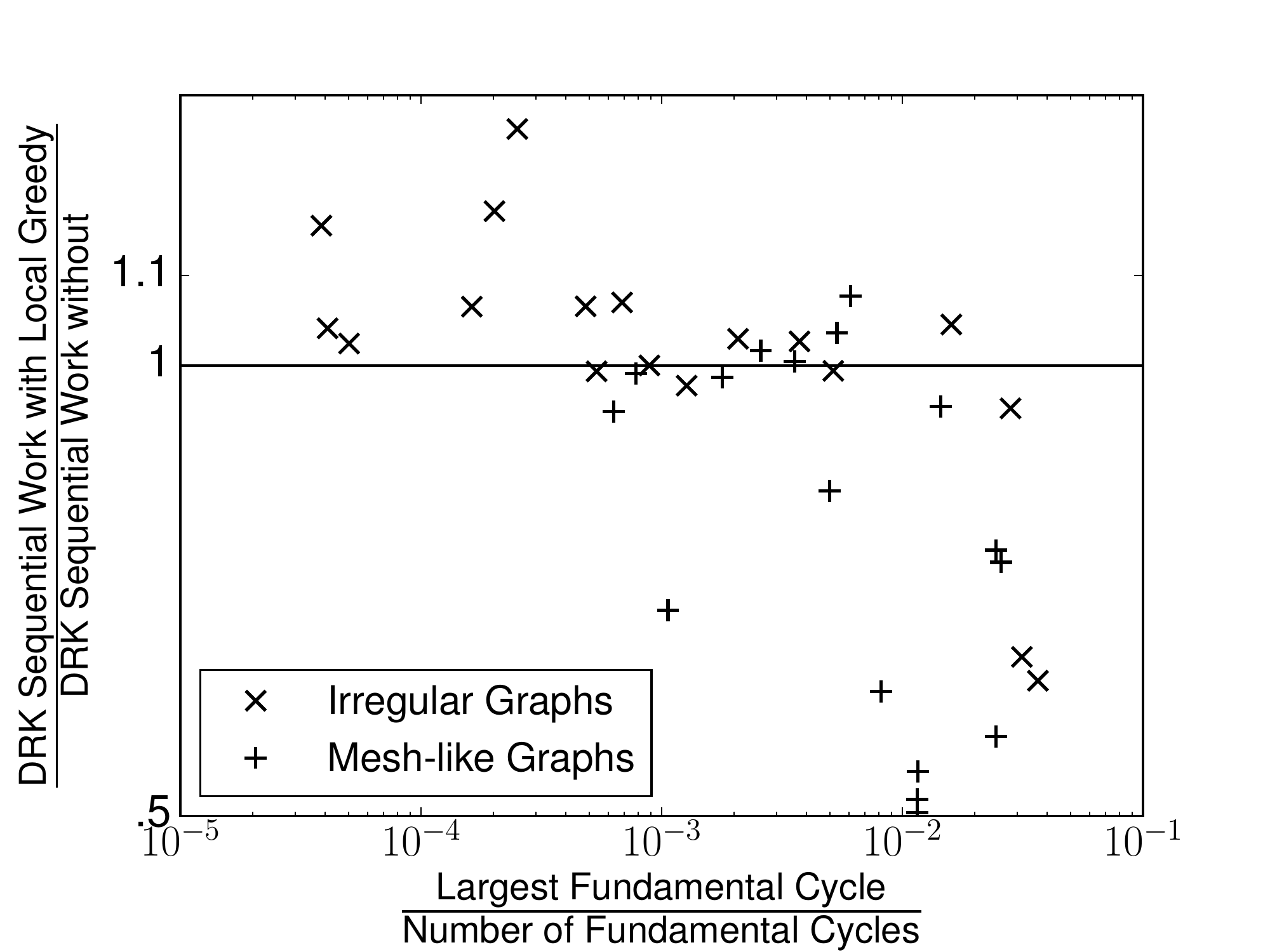}}
\caption{Sequential Comparison of Cycle Set Work: The ratio of DRK work with
and without local greedy cycles, on one thread,
is plotted against an estimate
of the usefulness of extra cycles.
Points below the line indicate that adding local greedy cycles helped.
\label{plot:cyclestats}}
\end{figure*}

We first examine the effects of using an expanded
cycle set in the sequential algorithm.
We estimate the usefulness of extra cycles
as the length of the largest cycle in the fundamental set normalized
by the number of cycles in the fundamental set. This is because
we suspect the large cycles to be a barrier to performance, as they
are updated the most frequently, and at the highest cost.
We plot the performance of the local greedy cycles
for the two different graph types, using metrics 1 and 4 in Figure
\ref{plot:cyclestats}).
These plots show the ratio between the work
of the expanded cycle sets as a function of the estimated
usefulness. Points below the line indicate that adding local greedy cycles
improved performance.

\begin{figure*}[htb]
\centering
\subfloat[2D Grid]{\includegraphics[scale=.35]{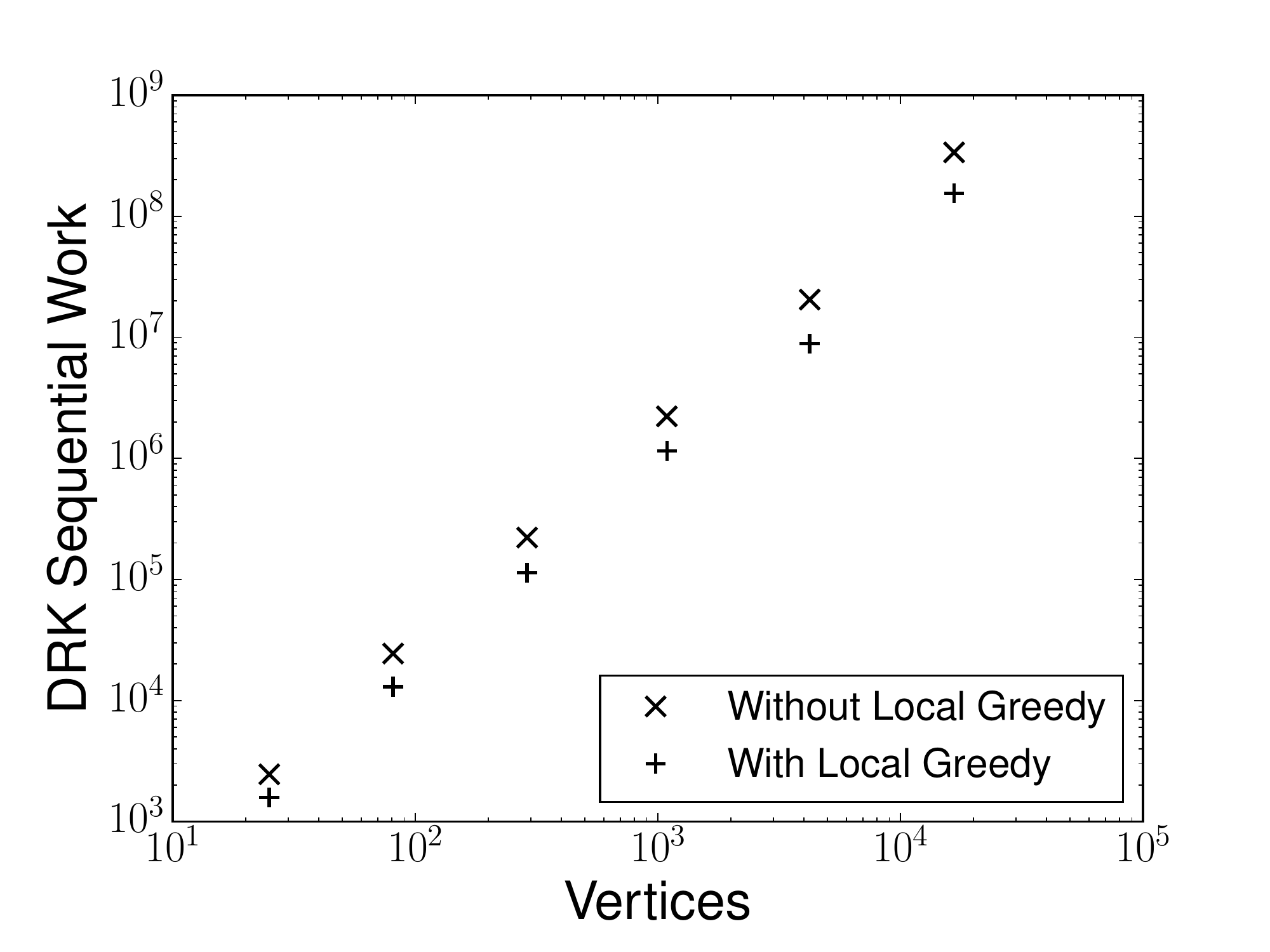}}
\subfloat[BTER]{\includegraphics[scale=.35]{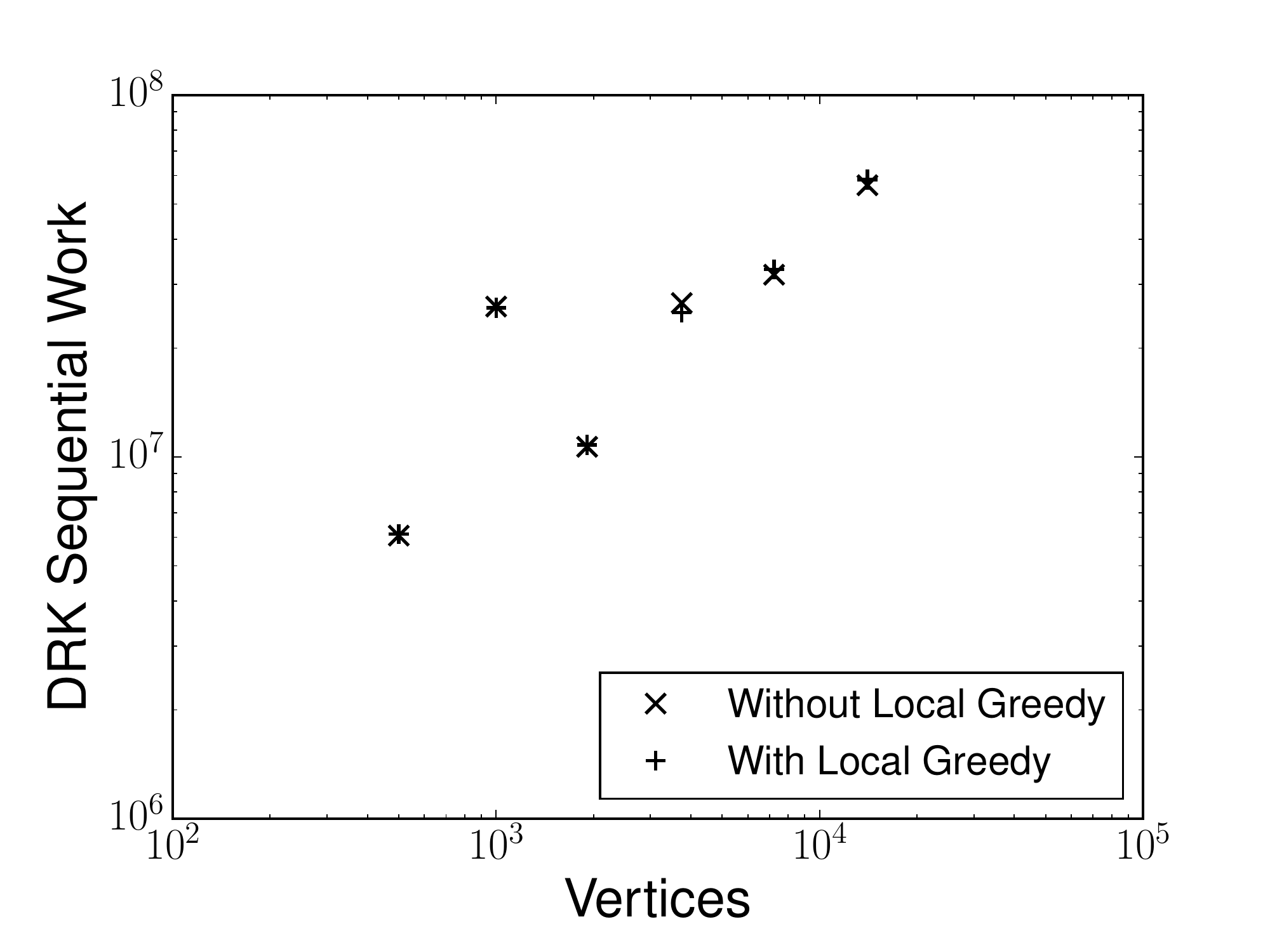}}
\caption{Weak Scaling of Cycle Set Work Under Cost Metric 1:
The DRK work with
and without local greedy cycles, on one thread,
is plotted against the graph size in vertices.
\label{plot:weakresults}}
\end{figure*}

We examine how the local greedy cycles perform as graph size increases
with weak scaling experiments on 2D grid graphs and BTER graphs.
The 2D grids used for this experiment are the same as in
Figure \ref{plot:gridresults}, and the BTER graphs were generated with
the parameters:
average degree of 20, maximum degree of $\sqrt{n}$,
global clustering coefficient
of 0.15, and maximum clustering coefficient of 0.15.
We plot the performance of cost Metric 1 as graph size scales
in Figure \ref{plot:weakresults}.

\begin{figure*}[htb!]
\centering
\subfloat[barth5]{\includegraphics[scale=.33]{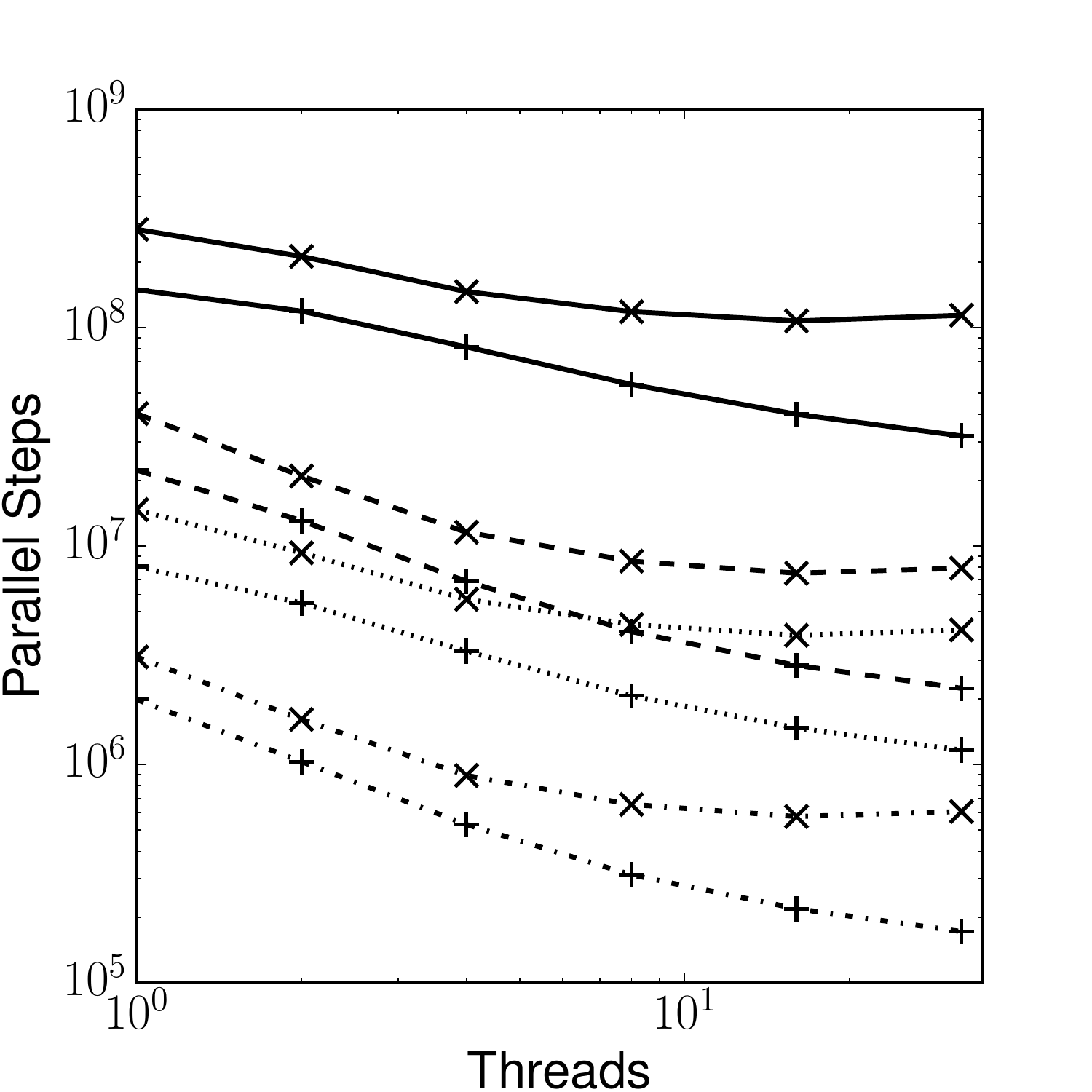}}
\subfloat[tuma1]{\includegraphics[scale=.33]{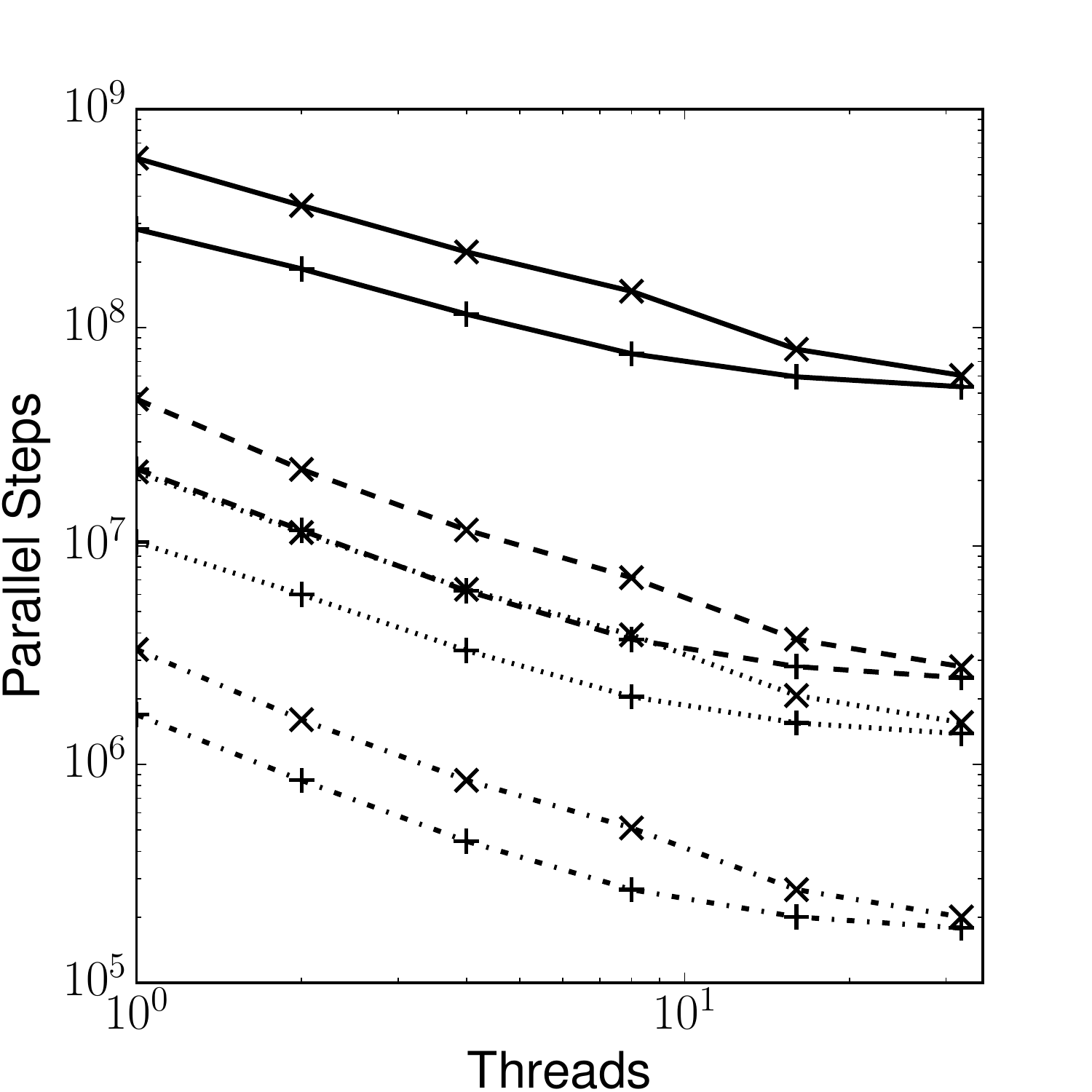}}
\captionsetup[subfigure]{oneside,margin={-1.3cm,0cm}}
\subfloat[email]{\includegraphics[scale=.33]{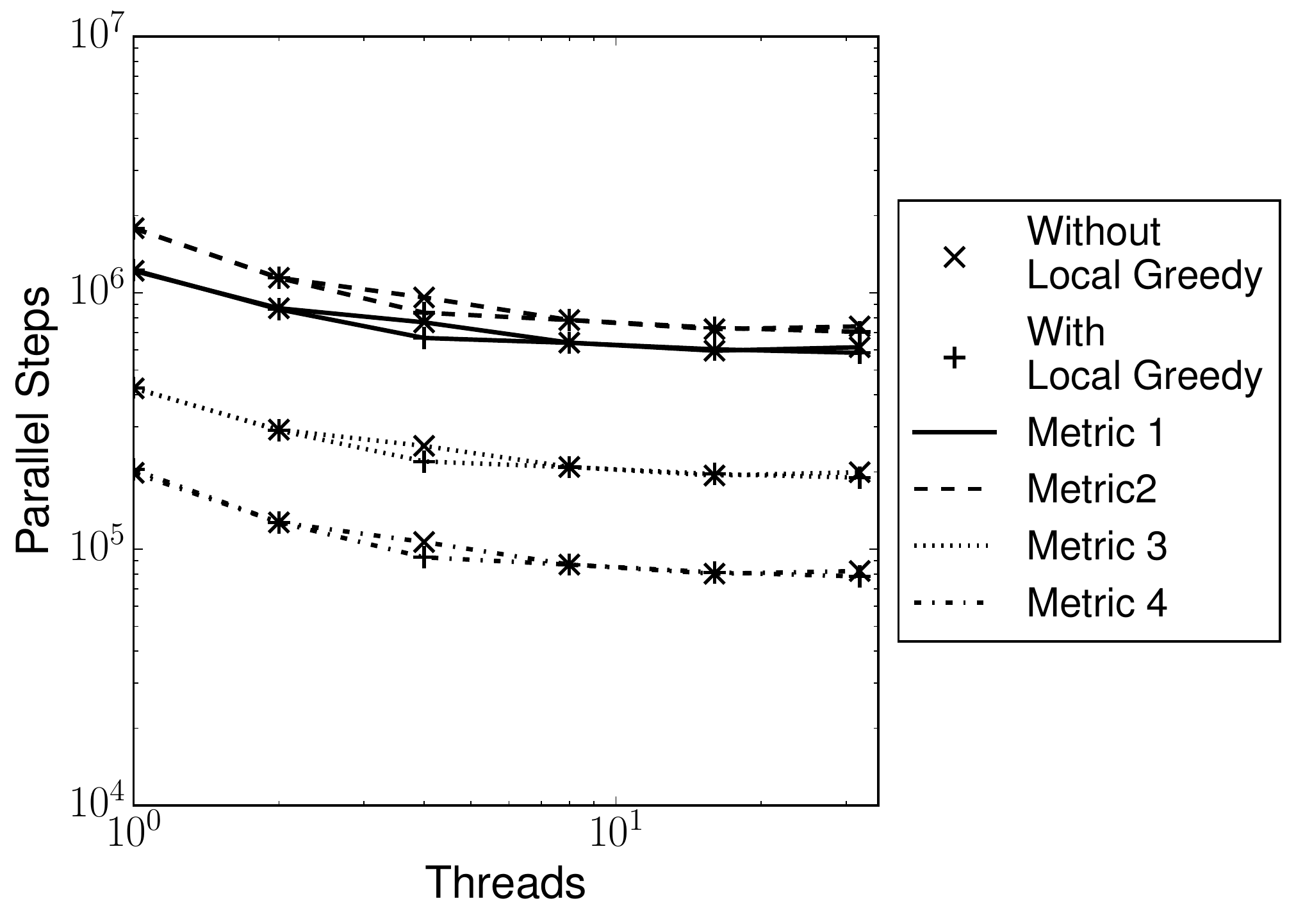}}
\caption{Parallel Steps Scaling (shown for three example graphs):
As threads are added, parallel steps decreases for both
cycle sets (steeper slope indicates
better scaling)\label{plot:example}}
\end{figure*}
\begin{figure*}[htb!]
\centering
\includegraphics[scale=.35]{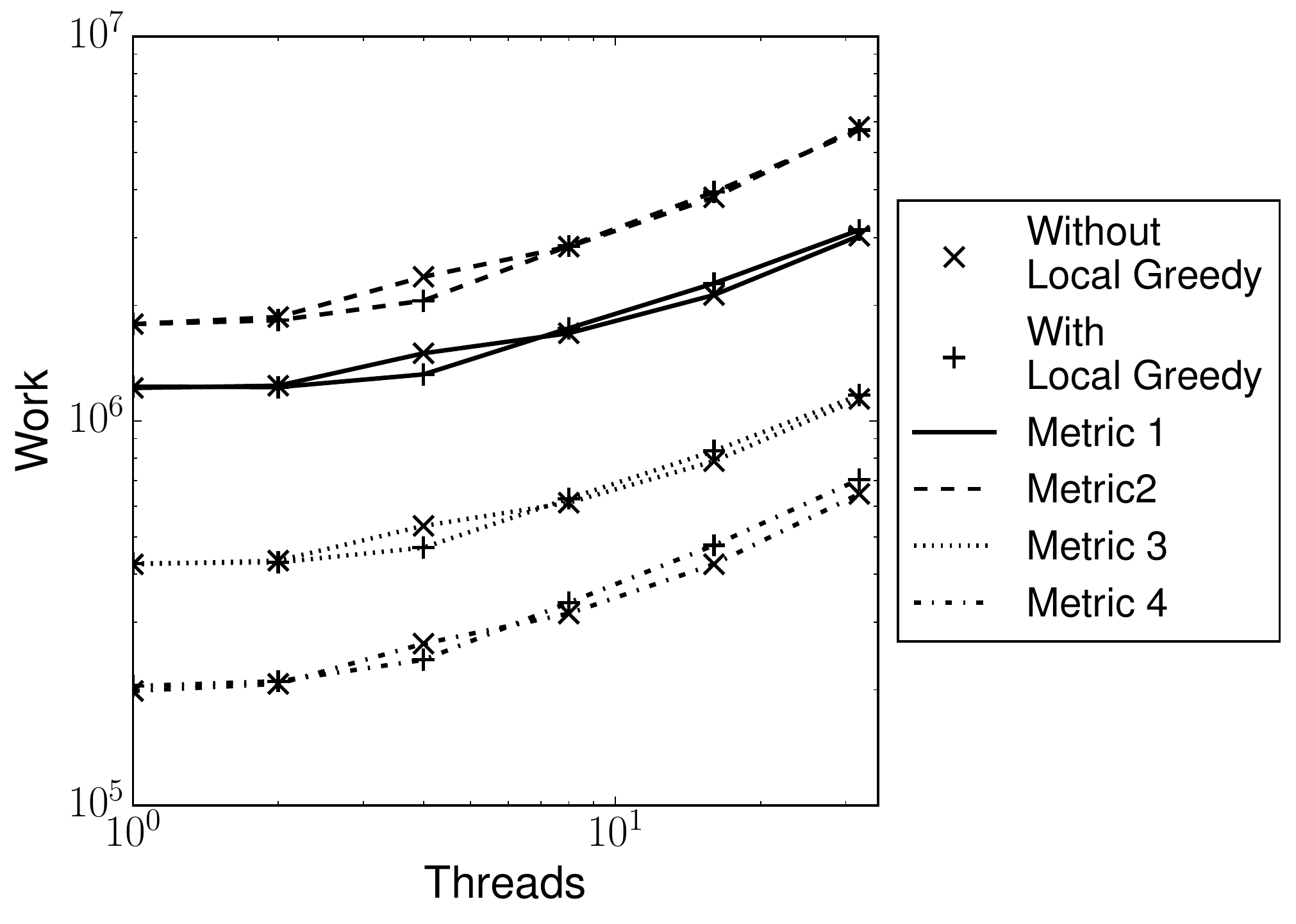}
\caption{Total Work Scaling of email Graph:
As threads are added the total
work increases for both cycle sets (ideally it would stay constant).}
\label{plot:exampleemailwork}
\end{figure*}
\begin{figure*}[htb!]
\centering
\subfloat[Mesh-like Graphs]{\includegraphics[scale=.35]{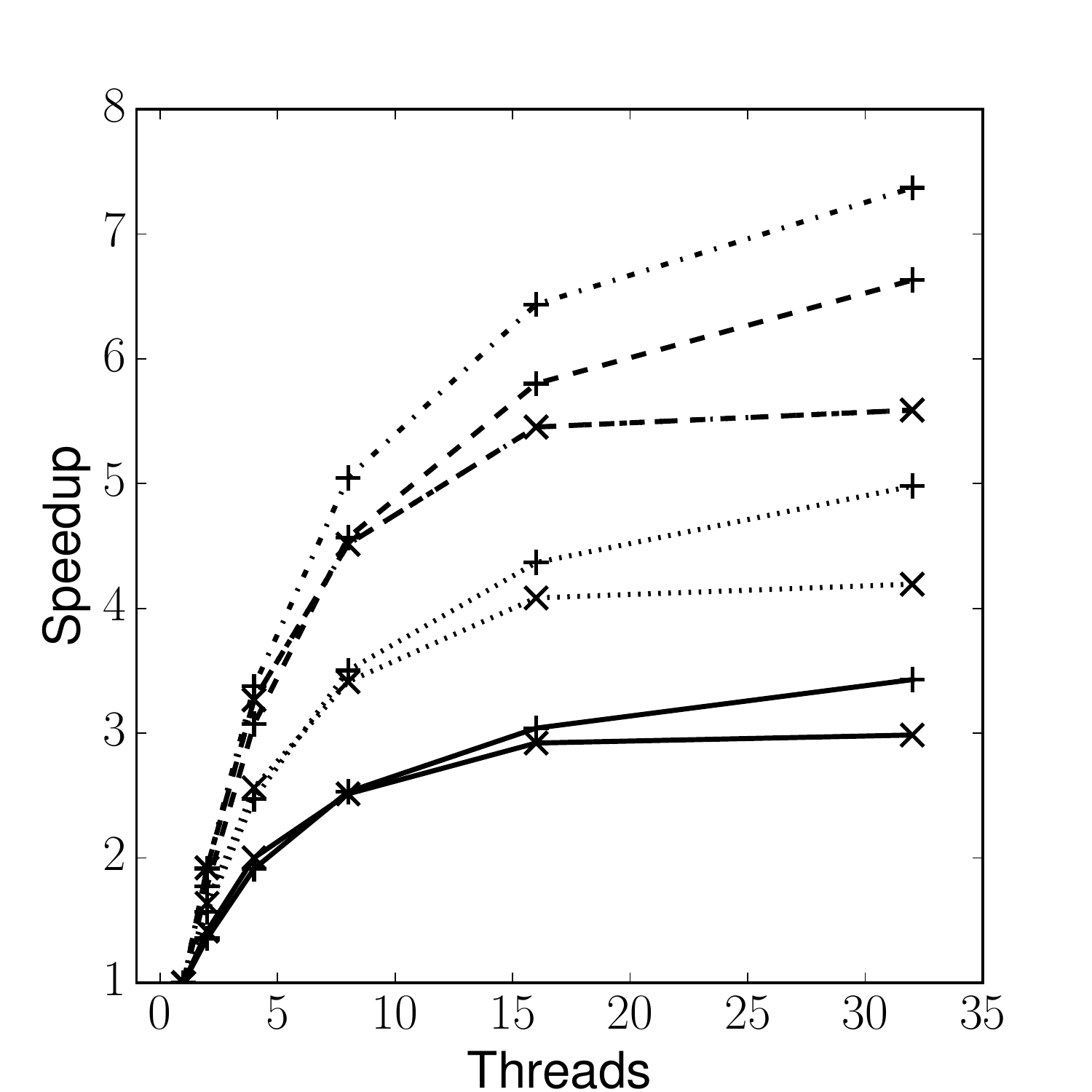}}
\subfloat[Irregular Graphs]{\includegraphics[scale=.35]{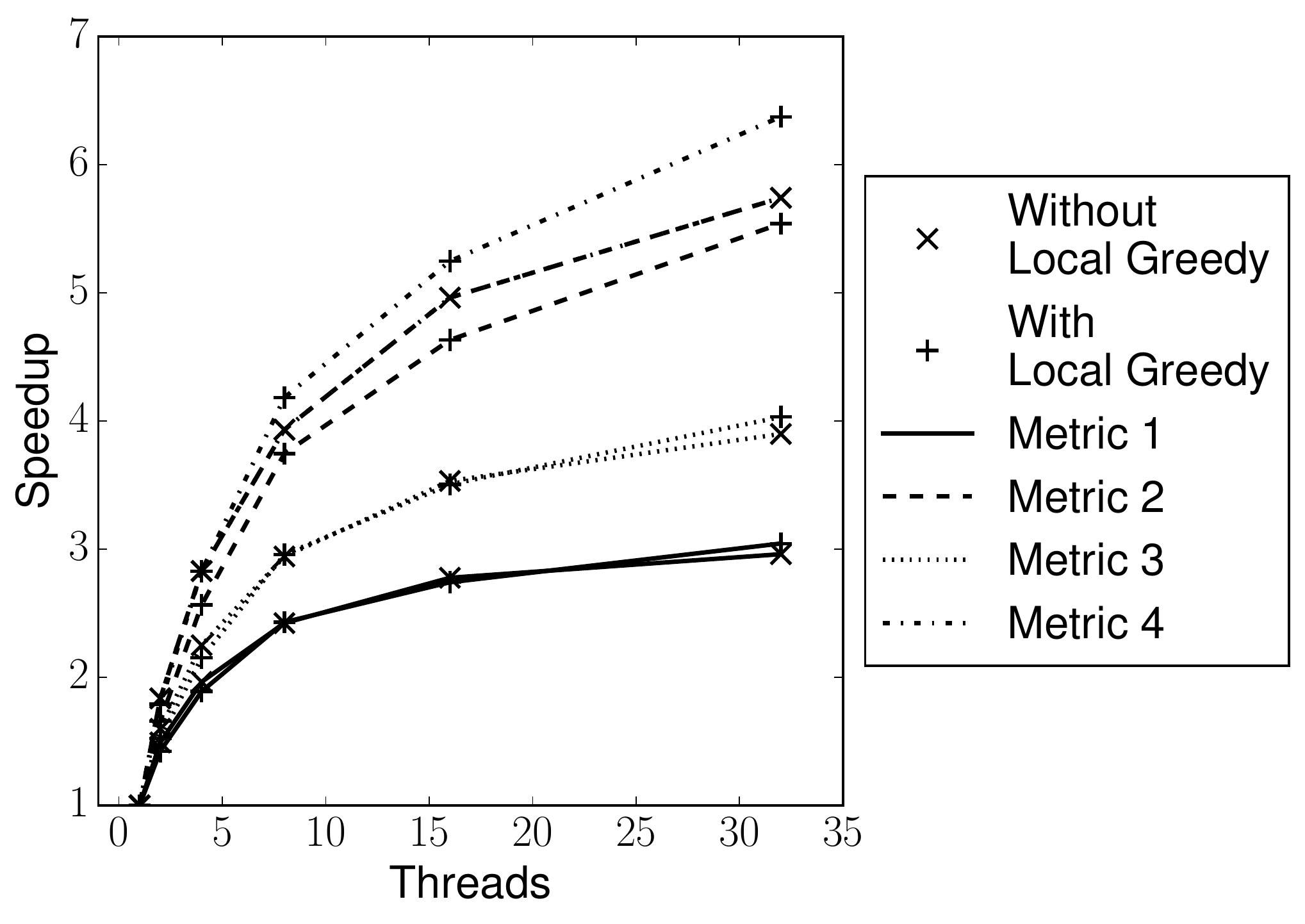}}
\caption{Average Parallel Steps Speedup: The ratio of sequential work on one thread to
parallel steps on multiple threads is plotted up to 32 threads.\label{plot:speedup}}
\end{figure*}
\begin{figure*}[htb!]
\centering
\includegraphics[scale=.35]{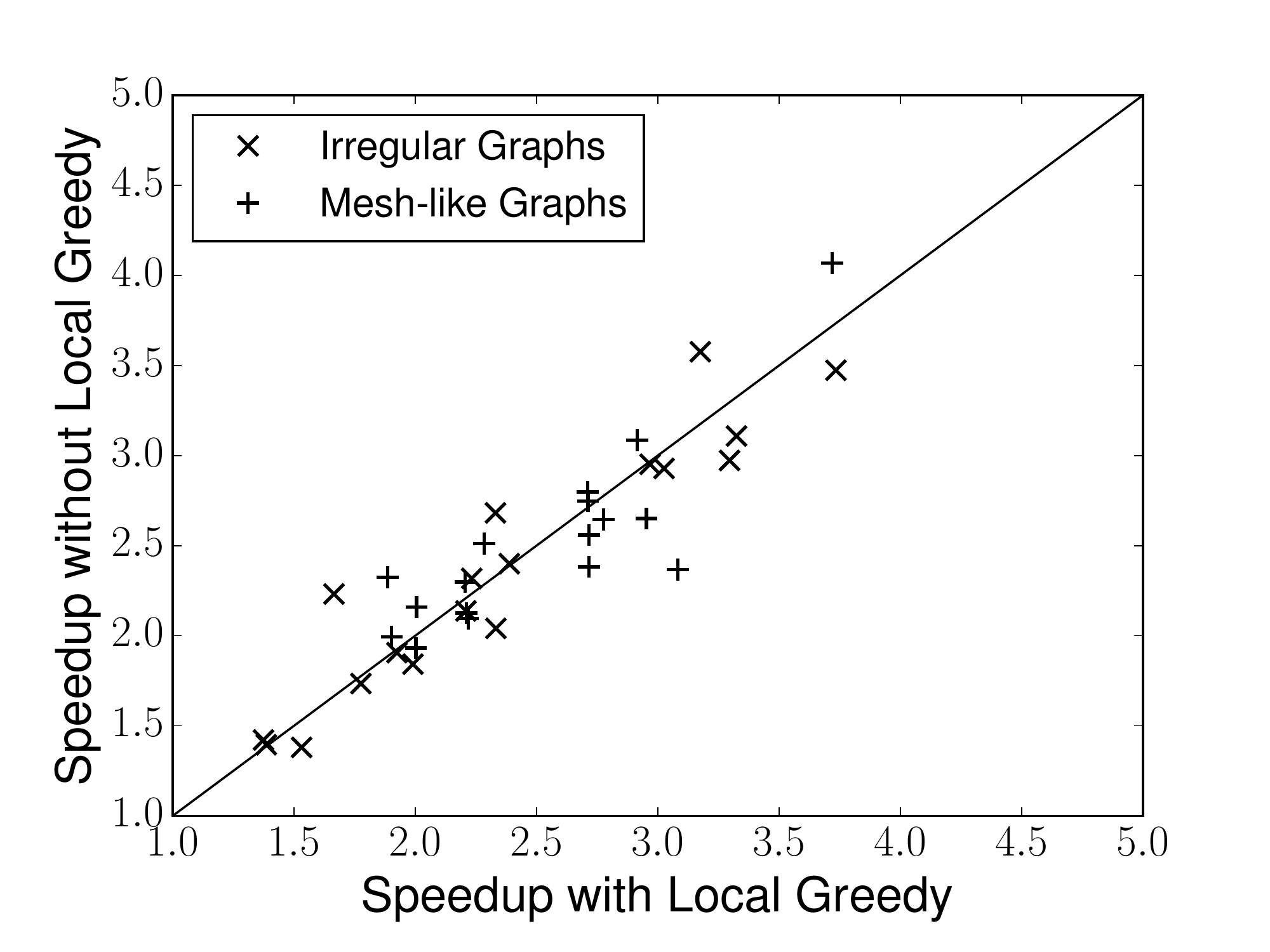}
\caption{8 Thread Speedup Comparison:
The ratio of the 8 thread speedups of both cycle sets are plotted
for all graphs (below the line local greedy speedup is better).}
\label{plot:speedup8}
\end{figure*}

Figure \ref{plot:example} shows
examples of our results on three of the graphs.
In Figure \ref{plot:example}(a)
the parallel steps
(with the four different metrics) is plotted as a
function of the number of threads
used for the barth5 graph. The total edges (metric 1) is at the
top of the plot, while the unit cost (metric 4) is at the bottom.
These results are shown for both fundamental and extended cycle sets.
Figure \ref{plot:exampleemailwork}
shows the effect of adding threads to the total work.

To measure the parallel performance across multiple graphs
we look at the average speedup of
the parallel steps across all graphs.
Speedup is defined as the sequential work using one thread over the
number of parallel steps using a number of multiple threads.
The speedup with and without extended cycles is shown in
Figure \ref{plot:speedup}. Note that without local greedy cycles
metric 2 and metric 4
speedup are the same as the costs differ by $\log{n}$.
We compare the speedup between the different cycle sets
for the different graph types in Figure \ref{plot:speedup8}.
Results are shown only for metric 1.
The speedup of using 8 threads without local greedy is plotted
against the speedup of using 8 threads with local greedy.

\subsection{Experimental Analysis}
In the sequential results shown in Figure \ref{plot:cyclestats},
there seems to be a threshold of largest cycle length above
which local greedy cycles can be useful,
but below which there is not much difference.
However, there is not a clear scaling with the cycle length ratio,
indicating that this is still a crude guess as to where
the extended cycle set is useful.
Also note that mesh-like graphs
tend to have larger girth (max cycle length) than
irregular graphs, leading to local greedy cycles working better on meshes.
The local greedy cycle improvement is slightly better for metric 1
where we count every edge update.
At the other extreme, when updating large cycles is the same cost (unit)
as small cycles added by local greedy, the local cycles are less effective.
However there is still an improvement in number of cycles updated.
We were unable to find some measure of
the usefulness of a single local greedy cycle.

The weak scaling experiments shown in Figure \ref{plot:weakresults}
show that, with the exception of a BTER outlier, the work scales nicely
with graph size. Also results for both cycle sets scale similarly.
The extended cycle set benefits the 2D grid graphs while
the BTER graphs see little improvement or are worse.
This is consistent with Figure \ref{plot:cyclestats} since the BTER
graphs are irregular and the 2D grids are mesh-like.

The scaling of parallel steps plots show
a variety of different behavior on the example graphs.
On the mesh-like barth5 graph
(shown in Figure \ref{plot:example}(a)),
the local greedy cycles improve both sequential
performance and the scaling of parallel steps performance.
At the left of this plot we see the extra cycles improve sequential results.
Then as threads are added in parallel, the steeper
slope indicates the local greedy cycles improved
the scaling of the parallel steps.
On the tuma1 graph
(shown in Figure \ref{plot:example}(b)),
the local greedy cycles improve sequential performance,
but result in similar or worse scaling.
At the left of this plot we see the extra cycles improve results
sequentially, but when scaled to 32 threads performance is similar.
On the email graph
(shown in Figure \ref{plot:example}(c)),
the local greedy cycles do not improve
sequential performance, and scaling is poor with both cycle sets.
There is little difference between the different cycle sets in this
plot. Furthermore scaling is poor and quickly flattens out by about
four threads.
For a better understanding of the poor parallel steps
scaling on the email graph, we examine the total work scaling
(shown in Figure \ref{plot:exampleemailwork}), showing how much extra
work we have to do when skewing the probability distribution.
This extra work quickly increases, limiting the parallel performance.

In the average parallel steps
speedup plot (shown in Figure \ref{plot:speedup}), we see
similar speedup for both cycle sets. On mesh-like graphs the local greedy
cycles do slightly better on all cost metrics beyond 16 threads. However
on the irregular graphs, only with cycle cost metric 4 do the local greedy
cycles perform better, and under metric 3 they perform worse. (Again note
that without local greedy cycles metric 4 and metric 3 speedups are the same.)
We hypothesized that giving the solver smaller, extra cycles
would improve the parallel
performance compared to the fundamental cycles.
However this seems to only be true for mesh-like graphs,
and even then the improvement is minimal.
An interesting thing to note is that the speedup
is better with the $\log{n}$ cost model. This is probably due to overcharging
small cycles, which is less problematic when there
are more threads to pick potentially
larger cycles.

Taking a snapshot of the parallel steps speedup results on eight threads
(shown in Figure \ref{plot:speedup8}),
we see that there are some irregular
graphs which do not have much speedup for either cycle set
(bottom left of the plot).
However there are mesh-like and irregular graphs
which enjoy a speedup for both cycle sets (top right
of the plot).
It is difficult to say on which graphs will different
cycles aid with parallelism.

\section{Conclusions}
We have done an initial comparison of Kelner et al.'s
DRK algorithm with PCG and PRK.
These preliminary results, measuring algorithm work by number of
edges touched or by number of cycles updated,
do not at present support the practical utility of DRK.
For mesh-like graphs, PCG usually takes less work than DRK, even if
DRK is charged only one unit of work per cycle update.
This suggests that the fast cycle update data structure proposed
by Kelner et al.\ (or any undiscovered fast update method) will not
be enough to make DRK practical.
It does seem that DRK is an improvement to PRK on several graphs,
mostly irregular graphs. One promising result of these experiments
is that DRK converges to small actual error similarly
to residual error, while PCG sometimes does not.
More PCG iterations are required
when solving to a low actual error, while DRK work does not
increase very much. More work should be done to
understand this behavior.

The experiments in this paper were limited to unweighted graphs
for simplicity.
Experiments with weighted graphs should be run for more
complete results.
An open question is whether there is a class of graphs with high
condition number, but with practical low stretch trees, where DRK
will perform significantly better in practice.

We suggest techniques for improving DRK in practice.
We consider using a spanning set, including non-fundamental
cycles, to accelerate
convergence. Using facial cycles of a two-dimensional grid graph
greatly reduces the required number of edge updates compared
to the fundamental cycle basis. We attempt to generalize these
cycles by finding small local greedy cycles. We found these
cycles can accelerate convergence, especially for mesh-like graphs.
It is difficult to
measure the usefulness of any one cycle in the basis, so it
is difficult to determine where and which extra cycles are
useful.

We also consider how DRK could be implemented in parallel to take
advantage of simultaneous updates of edge disjoint cycles. We
describe a model in which threads select cycles, and go idle
if a conflicting edge is found. While this can
increase total work, it can often
reduce the number of parallel steps.
However there is a limit to this parallelism.
Furthermore, this scaling behavior seems to be similar
with or without local greedy cycles.

\bibliographystyle{abbrv}
\bibliography{drk}
\begin{table*}[htb!]
\tiny
\centering
\subfloat[Mesh-like Graphs]{
\begin{tabular}{|c|c|c|c|c|c|c|c|}
\hline
Graph & Nodes & Edges & 2-core & 2-Core & Greedy & Probability of & Largest\\
(Collection) & & & Nodes & Edges & Cycles & Selecting Greedy & Cycle Length\\\hline
jagmesh3 (HB)& 1.09k & 3.14k & 1.09k & 3.14k & 1.92k & 0.2419 & 77\\\hline
lshp1270 (HB) & 1.27k & 3.70k & 1.27k & 3.70k & 2.17k & 0.4712 & 95\\\hline
rail\_1357 & 1.36k & 3.81k & 1.36k & 3.81k & 1.85k & 0.2507 & 55\\
(Oberwolfach) & & & & & & &\\\hline
50 x 50 grid & 2.50k & 4.90k & 2.50k & 4.90k & 2.40k & 0.5000 & 120\\\hline
data (DIMACS10)& 2.85k & 15.1k & 2.85k & 15.1k & 7.43k & 0.1760 & 92\\\hline
100 x 100 grid & 10.0k & 19.8k & 10.0k & 19.8k & 9.80k & 0.5000 & 230\\\hline
20 x 20 x 20 grid & 8.00k & 22.8k & 8.00k & 22.8k & 3.57k & 0.1941 & 122\\\hline
L-9 (A-G Monien) & 18.0k & 35.6k & 18.0k & 35.6k & 17.6k & 0.4992 & 411\\\hline
tuma1 & 23.0k & 37.2k & 22.2k & 36.5k & 10.7k & 0.0610 & 420\\
(GHS\_indef) & & & & & & & \\\hline
barth5 (Pothen)& 15.6k & 45.9k & 15.6k & 45.9k & 29.9k & 0.1765 & 375\\\hline
cti (DIMACS10)& 16.8k & 48.2k & 16.8k & 48.2k & 7.27k & 0.0501 & 172\\\hline
aft01 (Okunbor)& 8.21k & 58.7k & 8.21k & 58.7k & 26.6k & 0.6680 & 105\\\hline
30 x 30 x 30 grid & 27.0k & 78.3k & 27.0k & 78.3k & 8.35k & 0.1399 & 202\\\hline
wing (DIMACS10)& 62.0k & 122k & 62.0k & 122k & 27.9k & 0.0301 & 605\\\hline
olesnik0 & 88.3k & 342k & 88.3k & 342k & 220k & 0.1327 & 363\\
(GHS\_indef) & & & & & & &\\\hline
tube1 (TKK) & 21.5k & 438k & 21.5k & 438k & 0 & 0.0000 & 102\\\hline
fe\_tooth (DIMACS10) & 78.1k & 453k & 78.1k & 453k & 217k & 0.3673 & 286\\\hline
dawson5 (GHS\_indef) & 51.5k & 480k & 20.2k & 211k & 19.8k & 0.0941 & 165\\\hline
\end{tabular}}

\subfloat[Irregular Graphs]{
\begin{tabular}{|c|c|c|c|c|c|c|c|}
\hline
Graph & Nodes & Edges & 2-core & 2-core & Greedy & Probability of & Largest\\
(Collection) & & & Nodes & Edges & Cycles & Selecting Greedy & Cycle Length\\\hline
EVA (Pajek)& 8.50k & 6.71k & 314 & 492 & 84 & 0.2346 & 18\\\hline
bcspwr09 (HB)& 1.72k & 2.40k & 1.25k & 1.92k & 651 & 0.3276 & 54\\\hline
BTER1 & 981 & 4.85k & 940 & 4.82k & 510 & 0.0465 & 18\\
$d_{avg}=10$, $d_{max}=30$ & & & & & & &\\
$cc_{max}=.3$, $cc_{global}=.1$ & & & & & & &\\\hline
USpowerGrid (Pajek)& 4.94k & 6.59k & 3.35k & 5.01k & 1.68k & 0.2997 & 80\\\hline
email (Arenas)& 1.13k & 5.45k & 978 & 5.30k & 362 & 0.0433 & 11\\\hline
uk (DIMACS10)& 4.82k & 6.84k & 4.71k & 6.72k & 1.97k & 0.2488 & 211\\\hline
as-735 (SNAP) & 7.72k & 13.9k & 4.02k & 10.1k & 3.83k & 0.0822 & 9\\\hline
ca-GrQc (SNAP)& 4.16 & 13.4k & 3.41k & 12.7k & 4.43k & 0.2315 & 22\\\hline
BTER2 & 4.86k & 25.1k & 4.54k & 24.8k & 2.69k & 0.0468 & 17\\
$d_{avg}=10$, $d_{max}=70$ & & & & & & &\\
$cc_{max}=.3$, $cc_{global}=.1$ & & & & & & &\\\hline
gemat11 (HB)& 4.93k & 33.1k & 4.93k & 33.1k & 9.72k & 0.0011 & 42\\\hline
BTER3 & 4.94k & 37.5k & 4.66k & 37.2k & 4.79k & 0.0518 & 18\\
$d_{avg}=15$, $d_{max}=70$ & & & & & & &\\
$cc_{max}=.6$, $cc_{global}=.15$ & & & & & & &\\\hline
dictionary28 (Pajek)& 52.7k & 89.0k & 20.9k & 67.1k & 20.2k & 0.1410 & 36\\\hline
astro-ph (SNAP)& 16.7k & 121k & 11.6k & 111k & 13.2k & 0.0786 & 18\\\hline
cond-mat-2003 & 31.2k & 125k & 25.2k & 114k & 32.5k & 0.1533 & 23\\
(Newman) & & & & & & &\\\hline
BTER4 & 999 & 171k & 999 & 171k & 33 & 0.0002 & 7\\
$d_{avg}=15$, $d_{max}=30$ & & & & & & &\\
$cc_{max}=.6$, $cc_{global}=.15$ & & & & & & &\\\hline
HTC\_336\_4438 (IPSO)& 226k & 339k & 64.1k & 192k & 32.9k & 0.0339 & 990\\\hline
OPF\_10000 (IPSO)& 43.9k & 212k & 42.9k & 211k & 122k & 0.3146 & 53\\\hline
ga2010 (DIMACS10)& 291k & 709k & 282k & 699k & 315k & 0.1466 & 941\\\hline
coAuthorsDBLP & 299k & 978k & 255k & 934k & 297k & 0.1524 & 36\\
(DIMACS10) & & & & & & &\\\hline
citationCiteseer & 268k & 1.16M & 226k & 1.11M & 150k & 0.0484 & 56\\
(DIMACS10) & & & & & & &\\\hline
\end{tabular}}
\caption{Statistics of All Graphs Used in Experiments\label{table:graphsizes}}

\end{table*}

\begin{table*}[htb!]
\tiny
\centering
\subfloat[Mesh-like Graphs]{
\begin{tabular}{|c|c|c|c|}
\hline
Graph & Sequential Work & 2 Thread Parallel Steps & 8 Thread Parallel Steps \\
and Metric & (with Local Greedy) & (with Local Greedy) & (with Local Greedy) \\\hline
jagmesh3 (Metric 1) & 2.73M (1.72M) & 2.02M (1.26M) & 1.07M (28.3K)\\\hline
jagmesh3 (Metric 4) & 127K (101K) & 69.7K (51.2K) & 632K (17.4K)\\\hline
lshp1270 (Metric 1) & 6.80M (4.35M) & 4.87M (3.50M) &3.41M (2.29M) \\\hline
lshp1270 (Metric 4) & 192K (150K) & 104K (85.1K) & 61.0K (41.9K) \\\hline
rail\_1357 (Metric 1) & 1.64M (1.26M) & 1.21M (909K) & 653K (550K) \\\hline
rail\_1357 (Metric 4) & 119K (114K) & 63.8K (57.0K) & 143K (143K) \\\hline
50 x 50 grid (Metric 1) & 9.42M (4.32M) & 9.42M (4.43M) & 3.55M (1.50M) \\\hline
50 x 50 grid (Metric 4) & 213K (125K) & 115K (62.5K) & 45.0K (20.0K) \\\hline
data (Metric 1) & 17.9M (16.4M) & 13.1M (13.1M) & 8.43M (7.41M) \\\hline
data (Metric 4) & 815K (878K) & 416K (470K) & 185K (168K) \\\hline
100 x 100 grid (Metric 1) & 84.0M (38.1M) & 59.7M (29.1M) & 27.2M (13.1M) \\\hline
100 x 100 grid (Metric 4) & 1.11M (610K) & 560K (310K) & 180K (90.0K) \\\hline
20 x 20 x 20 grid (Metric 1) & 64.7M (63.3M) & 45.3M (46.0M) & 30.9M (28.6M) \\\hline
20 x 20 x 20 grid (Metric 4) & 1.55M (1.61M) & 816K (856K) & 448K (416K) \\\hline
L-9 (Metric 1) & 820M (382M) & 557M (266M)& 346M (124M) \\\hline
L-9 (Metric 4) & 3.92M (2.03M) & 2.07M (1.04M) & 1.10M (396K) \\\hline
tuma1 (Metric 1) & 597M (282M) & 362M (186M) & 147M (75.9M) \\\hline
tuma1 (Metric 4) & 3.36M (1.69M) & 1.60M (845K) & 512K (267K) \\\hline
barth5 (Metric 1) & 282M (149M) & 212M (119M) & 118M (54.9M) \\\hline
barth5 (Metric 4) & 3.11M (1.98M) & 1.61M (1.03M) & 655K (312K) \\\hline
cti (Metric 1) & 204M (195M) & 142M (143M) & 87.7M (103M) \\\hline
cti (Metric 4) & 3.87M (3.89M) & 2.02M (2.09M) & 1.01M (1.20M) \\\hline
aft01 (Metric 1) & 127M (118M) & 90.8M (88.3M) & 45.4M (43.5M) \\\hline
aft01 (Metric 4) & 4.38M (4.32M) & 2.19M (2.22M) & 763k (738k) \\\hline
30 x 30 x 30 grid (Metric 4) & 401M (394M) & 284M (283M) & 152M (142M) \\\hline
30 x 30 x 30 grid (Metric 4) & 6.51M (6.62M) & 3.35M (3.40M) & 1.35M (1.27M) \\\hline
wing (Metric 1) & 4.98B (4.16B) & 3.41B (2.99B) & 2.58B (2.08B) \\\hline
wing (Metric 4) & 21.8M (18.8M) & 12.2M (10.8M) & 8.31M (6.70M) \\\hline
olesnik0 (Metric 1) & 2.69B (1.71B) & 1.99B (1.36B) & 978M (630M) \\\hline
olesnik0 (Metric 4) & 33.8M (24.6M) & 16.9M (1.28M) & 5.47M (3.62M) \\\hline
tube1 (Metric 1) & 2.74B (N/A) & 1.98B (N/A) & 1.59B (N/A) \\\hline
tube1 (Metric 4) & 56.1M (N/A) & 32.0M (N/A) & 22.9M (N/A) \\\hline
fe\_tooth (Metric 1) & 6.56B (5.76B) & 4.46B (4.09B) & 3.04B (2.87B) \\\hline
fe\_tooth (Metric 4) & 65.5M (62.1M) & 34.3M (32.7M) & 19.8M (18.8M) \\\hline
dawson5 (Metric 1) & 1.49B (1.45B) & 1.06B (1.05B) & 650M (658M) \\\hline
dawson5 (Metric 4) & 24.9M (24.7M) & 12.8M (12.8M) & 6.11M (6.19M) \\\hline

\end{tabular}
}

\subfloat[Irregular Graphs]{
\begin{tabular}{|c|c|c|c|}
\hline
Graph & Sequential Work & 2 Thread Parallel Steps & 8 Thread Parallel Steps \\
and Metric & (with Local Greedy) & (with Local Greedy) & (with Local Greedy) \\\hline
EVA (Metric 1) & 41.4K (25.1K) & 22.5K (23.8K) & 18.5K (15.1K) \\\hline
EVA (Metric 4) & 6.28K (4.08K) & 2.83K (3.14K) & 1.88K (1.88K) \\\hline
bcspwr09 (Metric 1) & 308K (242K) & 199K (165K) & 130K (115K) \\\hline
bcspwr09 (Metric 4) & 26.2K (24.9K) & 12.5K (12.5K) & 4.99K (4.99K) \\\hline
BTER1 (Metric 1) & 2.26M (2.25M) & 1.78M (1.76M) & 1.59M (1.64M) \\\hline
BTER1 (Metric 4) & 246K (253K) & 243K (253K) & 333K (397K) \\\hline
USpowerGrid (Metric 1) & 1.06M (887K) & 735K (557K) & 297K (279K) \\\hline
USpowerGrid (Metric 4) & 73.8K (77.1K) & 36.9K (33.5K) & 10.1K (10.1K) \\\hline
email (Metric 1) & 1.22M (1.23M) & 871K (862K) & 639K (638K) \\\hline
email (Metric 4) & 199K (204K) & 127K (127K) & 87.0K (87.0K) \\\hline
uk (Metric 1) & 7.30M (3.88M) & 5.51M (3.29M) & 3.04M (1.62M) \\\hline
uk (Metric 4) & 160K (108K) & 84.8K (61.2K) & 33.0K (18.8K) \\\hline
as-735 (Metric 1) & 911K (887K) & 498K (550K) & 293K (267K) \\\hline
as-735 (Metric 4) & 201K (201K) & 96.6K (109K) & 48.3K (44.2K) \\\hline
ca-GrQc (Metric 1) & 2.98M (3.39M) & 1.92M (2.16M) & 923K (950K) \\\hline
ca-GrQc (Metric 4) & 413K (509K) & 201K (242K) & 71.7K (75.1K) \\\hline
BTER2 (Metric 1) & 11.3M (11.7M) & 8.21M (8.31M) & 6.52M (6.60M) \\\hline
BTER2 (Metric 4) & 1.30M (1.39M) & 876K (894K) & 658K (667K) \\\hline
gemat11 (Metric 1) & 63.7M (62.2M) & 50.4M (49.4M) & 45.7M (44.9M) \\\hline
gemat11 (Metric 4) & 3.30M (3.22M) & 2.38M (2.33M) & 2.07M (2.04M) \\\hline
BTER3 (Metric 1) & 18.5M (19.0M) & 14.6M (14.1M) & 13.4M (12.4M) \\\hline
BTER3 (Metric 4) & 2.13M (2.27M) & 1.59M (1.54M) & 1.39M (1.29M) \\\hline
dictionary28 (Metric 1) & 38.0M (33.8M) & 24.4M (24.6M) & 16.4M (15.2M) \\\hline
dictionary28 (Metric 4) & 3.20M (3.18M) & 1.65M (1.78M) & 941K (878K) \\\hline
astro-ph (Metric 1) & 25.3M (25.9M) & 15.7M (16.3M) & 8.63M (8.57M) \\\hline
astro-ph (Metric 4) & 4.45M (4.74M) & 2.27M (2.43M) & 1.01M (1.01M) \\\hline
cond-mat-2003 (Metric 1) & 38.8M (40.9M) & 27.4M (30.3M) & 21.0M (20.5M) \\\hline
cond-mat-2003 (Metric 4) & 4.57M (5.35M) & 2.62M (3.10M) & 1.74M (1.72M) \\\hline
BTER4 (Metric 1) & 25.4M (26.4M) & 15.1M (15.1M) & 8.53M (8.01M) \\\hline
BTER4 (Metric 4) & 6.78M (7.06M) & 3.67M (3.67M) & 1.84M (1.73M) \\\hline
HTC\_336\_4438 (Metric 1)& 14.2B (13.7B) & 8.50B (8.89B) & 4.10B (3.67B) \\\hline
HTC\_336\_4438 (Metric 4)& 32.2M (32.0M) & 15.9M (16.9M) & 6.80M (6.09M) \\\hline
OPF\_10000 (Metric 1) & 47.3M (49.1M) & 31.8M (33.2M) & 16.0M (16.5M) \\\hline
OPF\_10000 (Metric 4) & 6.86M (8.66M) & 3.34M (4.29M) & 857K (1.07M) \\\hline
ga2010 (Metric 1) & 8.77B (6.16B) & 6.88B (4.97B) & 3.21B (2.59B)\\\hline
ga2010 (Metric 4) & 40.8M (33.5M) & 20.8M (16.9M) & 6.48M (5.35B)\\\hline
coAuthorsDBLP (Metric 1) & 539M (552M) & 356M (371M) & 264M (237M) \\\hline
coAuthorsDBLP (Metric 4) & 44.4M (51.3M) & 23.5M (26.3M) & 15.3M (13.8M) \\\hline
citationCiteseer (Metric 1) & 1.08B (1.07B) & 716M (723M) & 506M (482M) \\\hline
citationCiteseer (Metric 4) & 74.4M (76.2M) & 40.7M (41.8M) & 25.3M (24.2M) \\\hline

\end{tabular}
}
\caption{Condensed Results of Scaling Experiments\label{table:results}}
\end{table*}

\end{document}